\documentclass[pre,reprint,floatfix,showpacs,showkeys]{revtex4-1}

\usepackage{graphicx}
\usepackage{amsmath}

\begin{document}

\title{Instabilities and mixing in two-dimensional Kolmogorov flow}

\author{Radford Mitchell, Jr.}
\author{Roman O. Grigoriev}
\affiliation{Center for Nonlinear Science and School of Physics, Georgia Institute of Technology, 837 State Street, Atlanta, GA 30332-0430}

\date{\today}

\begin{abstract}
This paper presents results of a theoretical investigation of transport in a numerical model of a two-dimensional Kolmogorov flow. We investigate the changes in its mixing properties associated with transition from laminar regime to turbulence. It is found that significant changes in the flow do not always lead to comparable changes in its transport properties. On the other hand, some very subtle changes in the flow can dramatically alter the degree of mixing. We show that interaction of multiple resonances can provide an explanation for many of these seemingly paradoxical results.
\end{abstract}

\maketitle

\section{Introduction}
Two-dimensional (2D) flows have proven to be very useful for studying various phenomena in fluid dynamics since, compared to three-dimensional flows, they are much more amenable to theoretical analysis and numerical investigation. In particular, much of our fundamental understanding of transport properties of fluid flows has been developed using 2D models. Effectively 2D fluid flows are responsible for transport and mixing in many geophysical processes such as atmospheric \cite{Haynes2005,Pierrehumbert1991} and oceanic \cite{Haller1997,Wiggins2005} flows as well as in convection processes within the Earth's mantle \cite{Hoffman1985,Allegre1986}. Two-dimensional laminar mixing is a key process in many types of microfluidic essays \cite{Lee2001,Niu2003}, such as ones used for gene expression profiling \cite{Stremler2004}, and in numerous technological applications, such as the production of polymer blends \cite{Ottino2004}. The reduction to two dimensions has also provided insights into many difficult 3D problems ranging from mixing in the radiation zones of rotating stars \cite{Mathis2004} to confinement of thermonuclear plasmas \cite{Robert1980}. 

Much of our understanding of transport properties of fluid flows comes from experimental observations or numerical simulations of the advection, or stirring, of passive tracers by the flow. The dynamics of passive tracers in 2D flows of incompressible fluids is formally described by a Hamiltonian system with one degree of freedom
\begin{eqnarray}
\label{tracers}
&&\dot{x}=v_x=\partial_y\Psi,\nonumber\\
&&\dot{y}=v_y=-\partial_x\Psi,
\end{eqnarray}
where the stream function $\Psi(x,y,t)$ plays the role of a Hamiltonian and the coordinates $x$ and $y$ are the conjugate variables.

Time-independent 2D flows are integrable (because $\Psi$ is conserved), and the trajectories of tracers coincide with the closed stream lines of the flow, resulting in poor mixing. The introduction of time-dependence effectively makes the velocity field a three-dimensional flow, which is generally nonintegrable. Stream lines in such flows can be chaotic even if the underlying velocity field is regular (e.g., stable and time-periodic). If this is the case, the stream lines will diverge exponentially fast from one another, resulting in rapid stretching and folding of fluid elements. This process, known as chaotic advection or Lagrangian chaos, is the underlying mechanism responsible for dramatically improved mixing.
 
One of the simplest 2D models in which chaotic mixing can occur is the `blinking-vortex' flow studied by Aref \cite{Aref1984}. Historically the first study of the mixing properties of a fluid system, this model was originally proposed as an idealization of a periodically stirred fluid and consists of a pair of spatially separated fixed point vortices which are alternately turned on for one half of the period $T$. Numerical simulations showed that when both vortices are running continuously (i.e., $T=0$), the flow is time-independent and, thus, integrable. For small nonzero $T$, it was found that the trajectories nearest the vortices become chaotic. The size of the mixed region increases monotonically with $T$ until, at some finite critical value of $T$, the entire domain becomes uniformly mixed.
 
The first analytic investigation of mixing in a time-periodic 2D flow is due to Khakhar \textit{et al}. \cite{Khakhar1986}. This study introduced an idealized model, known as the `tendril-whorl' flow, in which uniform shear is followed by differential rotation and showed that mixing takes place in the vicinity of separatrices associated with saddle fixed points of the period-$T$ map of the flow, while the KAM tori surrounding the elliptic fixed points serve as transport barriers. The same structures were shown to also control mixing in the `blinking vortex' flow. These studies demonstrated that laminar, time-periodic, area-preserving 2D flows can produce efficient mixing.
 
The results of these idealized models raised questions as to whether or not real-world laminar fluid flows could give rise to chaotic stream lines. This prompted the analytical and numerical study of chaotic advection in a journal bearing Stokes flow with physical boundary conditions \cite{Aref1986,Chaiken1987}. The basic setup is that of a Couette flow between non-coaxial rotating cylinders, where time-periodicity is introduced by alternating the rotation between the inner and outer cylinders. By varying the distance between the axes of the cylinders as well as the time-interval for which one of the cylinders rotates, one can obtain various flow patterns with both regular and chaotic trajectories. The experimental realization of this flow \cite{Chaiken1986} showed excellent agreement with numerical results. Subsequently, the experimental study of cavity flows by Chien {\em et al.} \cite{Chien1986} showed the existence of transverse intersections of homo/heteroclinic manifolds at small Reynolds numbers, providing more evidence for the mixing capabilities of 2D laminar flows. 
 
Rom-Kedar \textit{et al}. \cite{Rom-Kedar1990} proved the existence of chaotic trajectories analytically for a model flow produced by a pair of time-independent point vortices perturbed by a time-periodic shear. The theory of lobe dynamics developed in this paper set up a framework for quantitative description of transport across separatrices of the unperturbed flow which evolve into a homoclinic tangle in the presence of perturbation. In conjunction with the analytic techniques introduced by Melnikov \cite{melnikov1963}, this framework enabled them to estimate the fluxes between different regions of the flow domain.

Recently, most experimental and many theoretical studies of mixing in 2D flows have used thin layers of electrolyte placed over various arrangements of permanent magnets. The fluid flow is driven by the Lorenz force which arises when electric current flows through the electrolyte. Since this setup is closely related to our work, we describe here other studies which used it.

Rothstein {\em et al.} \cite{Rothstein1999} discovered the existence of persistent spatial patterns, which they called strange eigenmodes, in a flow driven by a combination of time-periodic current and either a disordered or a square array of magnets. These patterns were shown to emerge as a result of a delicate balance between advective stretching and molecular diffusion. The process of mixing was observed to continue even after these structures reached an asymptotic shape. The same experimental setup was subsequently used to investigate the rate of mixing \cite{Voth2003}. By examining the spatial structure of persistent patterns, it was found that locally, mixing rates are controlled by stretching, but on large scales they are governed by diffusive transport. Additionally, it was discovered that mixing rates could be dramatically enhanced by breaking certain spatial and temporal symmetries.

Voth {\em et al.} \cite{Voth2002} used a disordered array of magnets and time-periodic current to drive the flow and were able to use flow measurements to construct forward and backwards finite-time Lyapunov exponent (FTLE) fields which follow the time-evolution of the unstable and stable manifolds of saddle points of the flow, thus providing an empirical method for visualizing the geometrical structures underlying the mixing process. A follow-up experimental study carried out using magnets arranged in a square, hexagonal, and a disordered array \cite{Arratia2005} found that the probability distribution of FTLEs exhibited self-similar behavior regardless of the flow pattern or the degree of mixing in the system.

Fluid mixing was also studied in time-dependent flows driven by steady current. Danilov {\em et al.} \cite{Danilov2000} performed a combined experimental and theoretical study of mixing by a time-periodic four-vortex flow. Numerical study of a truncated analytic model showed that separatrices partitioned the flow domain into regions with different mixing rates and that transport between these regions was a relatively slow process compared to the mixing within these regions. The theory of adiabatic chaos was used to explain the results and show that long-term transport could effectively be modeled as a random walk of an adiabatic invariant.  

This paper investigates mixing properties of a range of 2D flows arising as intermediate stages in the transition from the so-called Kolmogorov flow \cite{Arnold1960,Meshalkin1961} to turbulence. Unlike the majority of other studies of mixing in 2D flows where the time-dependence is due to external monochromatic forcing, our focus is on time-dependence that arises naturally as a result of fluid-dynamic instabilities, producing flows ranging in their temporal complexity from time-periodic to quasi-periodic and chaotic. The paper is organized as follows: In Sect. \ref{s:model} we introduce the model of the fluid flow and characterize the flow states that emerge in the transition from the laminar to the turbulent regime. The mixing properties of these flows are described and analyzed in Sect. \ref{s:mixing}. Finally, summary and conclusions are presented in Sect. \ref{s:summary}.

\section{Problem Description}
\label{s:model}

\subsection{Model of the Fluid Flow}

We consider a model of an experimental flow described in Ref. \onlinecite{jfm_12} which employs bar magnets with alternating polarity to generate a Kolmogorov flow in a layer of electrolyte supported by a liquid dielectric. 
The flow in the conducting layer can be described by the following equation for the vorticity $\Omega=-\nabla^2\Psi$:
\begin{equation} 
\partial_t\Omega+\beta{\bf v}\cdot\nabla\Omega=\nu\nabla^2\Omega-\alpha\Omega+A\sin ky 
\label{vorticity}
\end{equation}
where $k=\pi/w$ and $w$ is the width of individual magnets. Parameters $\beta=1$, $\nu=0.0115$ cm$^{2}$/s, and $\alpha=0.1141$ s$^{-1}$ were selected to be representative of a typical experimental setup. Furthermore, we chose the domain width $L_y=5$ cm corresponding to four magnets of width $w=1.25$ cm and the length $L_x=2L_y=10$ cm. For simplicity, unlike the experimental system which is larger and features physical (no-slip) lateral boundary conditions, we assume periodic boundary conditions. The effect of the bottom boundary, however, is included in our model via the Rayleigh friction term $-\alpha \Omega$. The importance of this term is described by the non-dimensional combination $F=\alpha/\nu k^2\approx 1.57$ which shows that it is comparable to the viscous term $\nu\nabla^2\Omega$. Finally, $A$ measures the strength of the driving force and is used as a control parameter analogous to the Reynolds number.

The vorticity equation (\ref{vorticity}) was solved numerically using a spectral (Fourier) method with $64 \times 128$ modes. As a check, we recalculated the bifurcation sequence using $128 \times 256$ modes which yielded less than a $1\%$ difference in both the leading stability eigenvalues and the location of the bifurcations. Temporal discretization used a second-order, implicit-explicit, operator-splitting scheme with an adaptive time step \cite{Ascher1995}. The Crank-Nicolson method was used for the linear and forcing terms in Fourier space, while the fourth-order Runge-Kutta method was used for the advection term in real space. The use of the so-called Strang-Marchuk splitting \cite{Marchuk1975} ensures that the resulting scheme is second order in time. 

\subsection{From Kolmogorov Flow to Turbulence}

In this section we describe the transition to 2D turbulence in our model system as the value of the control parameter $A$ is increased. Unlike many shear flows in 3D which transition directly from laminar flow to turbulence, here we find a rather complicated sequence of transitional flow states whose temporal complexity changes in a rather non-monotonic fashion before a turbulent flow is eventually established.

\begin{figure}
(a)\includegraphics[width=3in]{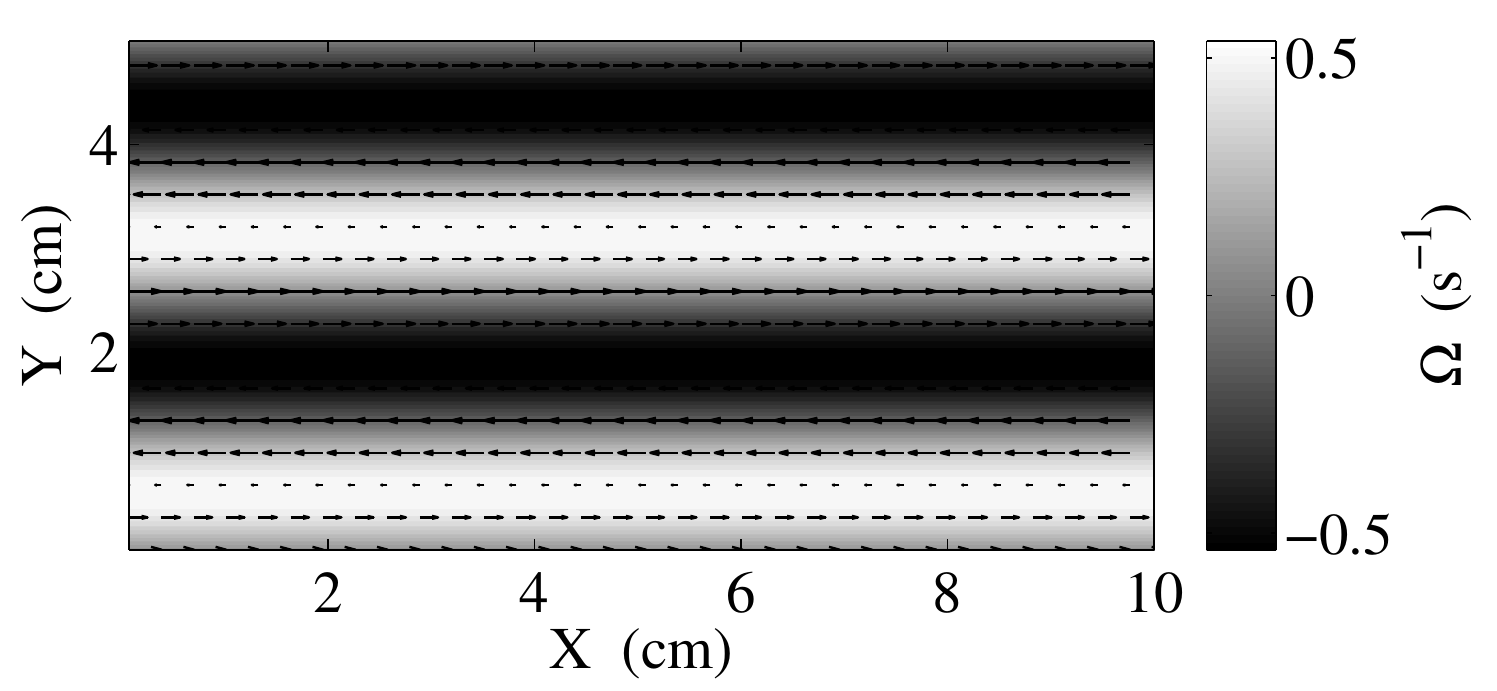}\\
(b)\includegraphics[width=3in]{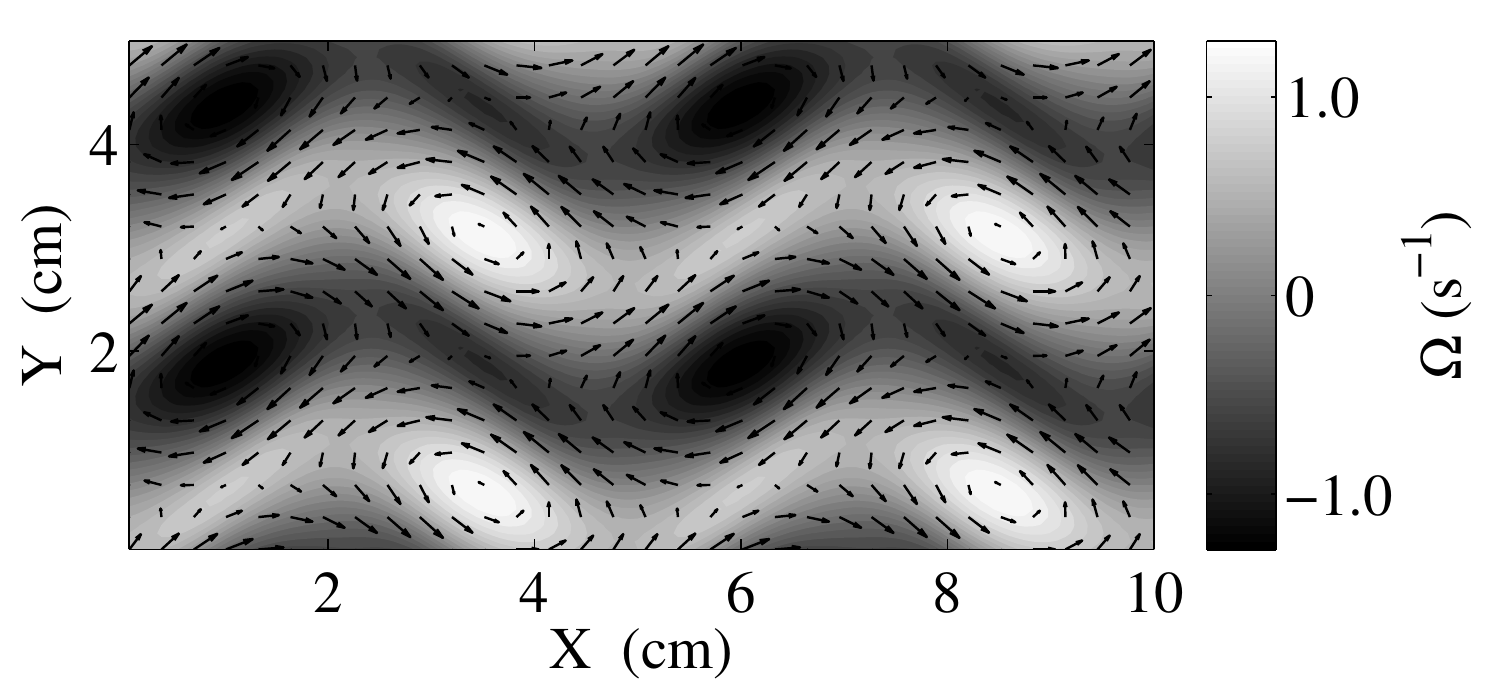}
\caption{Laminar flow $L$ at $A=0.1$ s$^{-2}$ (a) and spatially modulated flow $M$ at $A=0.250$ s$^{-2}$ (b).
Velocity field (arrows) is overlayed on top of vorticity field (grayscale).}
\label{LM}
\end{figure}

\begin{figure}
\includegraphics[width=3in]{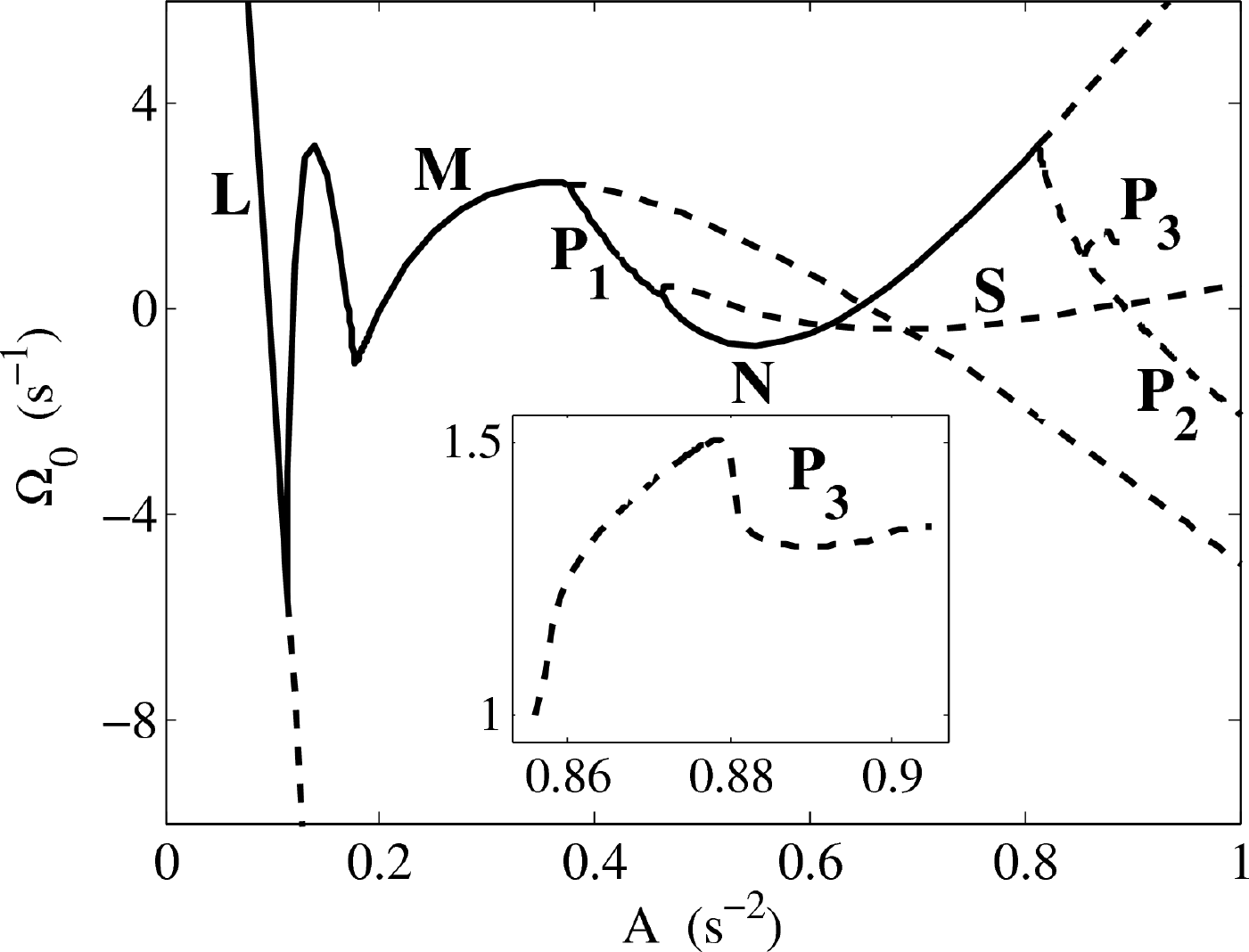}
\caption{Bifurcation diagram. The relative vorticity magnitude $\Omega_0\equiv \|\Omega-\Omega_L\|_2-cA$ is shown, where $c$ is a constant chosen to separate the various branches of the diagram for visualization purposes. Solid and dashed lines denote stable and unstable states, respectively. Periodic orbits are represented by their time-averaged values. Inset shows the region where the $P_3$ branch exists.}
\label{Bifurcation}
\end{figure}

Kolmogorov flow profile describes a laminar solution of the vorticity equation (\ref{vorticity}) with the symmetry of the driving: continuous translational symmetry in the $x$ direction and discrete translational symmetry in the $y$ direction. The problem also possesses two additional discrete symmetries (rotation by 180 degrees about  a vertical axis and a flip about $x$ (or $y$) axis combined with the change in the sign of vorticity), but these will not play an important role in the subsequent discussion. The laminar flow (referred to simply as $L$ below) is described by the following analytical solution for the vorticity 
\begin{equation} 
\Omega_L=\frac{A}{\alpha + k^2 \nu}\sin ky
\label{laminar}
\end{equation}
and features straight alternating shear bands which reflect the geometric arrangement of the magnets (see Fig. \ref{LM}(a)). For our choice of parameters, linear stability analysis predicts this flow profile to be stable for $A<0.1145$ s$^{-2}$. This is confirmed by the results of our numerical simulations summarized in Fig. \ref{Bifurcation}, which shows all stable and unstable solutions that have been computed using a Jacobian-free Newton-Krylov solver \cite{Knoll2004} for $A\leq 1$ s$^{-2}$.

At $A\approx 0.1145$ s$^{-2}$ the laminar flow $L$ loses stability through a supercritical pitchfork bifurcation and is replaced with its steady, spatially modulated version. As $A$ is increased, the distortion of the shear bands increases and they are gradually replaced with a periodic array of counter-rotating vortices. This spatially modulated shear flow (denoted $M$ and shown in Fig. \ref{LM}(b)) eventually undergoes a supercritical Hopf bifurcation and loses stability at $A \approx 0.3750$ s$^{-2}$. 

\begin{figure}[b]
\includegraphics[width=\columnwidth]{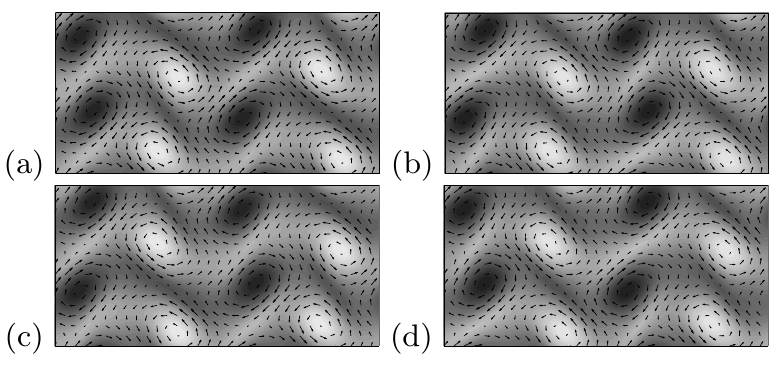}
\caption{Time-periodic flow $P_1$ at $A=0.428$ s$^{-2}$ and (a) t=0, (b) t=T/4, (c) t=T/2, (d) t=3T/4 with $T=365.83$ s. The same color bar as in Fig. \ref{NS}(a) is used here.}
\label{P1}
\end{figure}

\begin{figure}[t]
\centering
(a)
\includegraphics[height=1.15in]{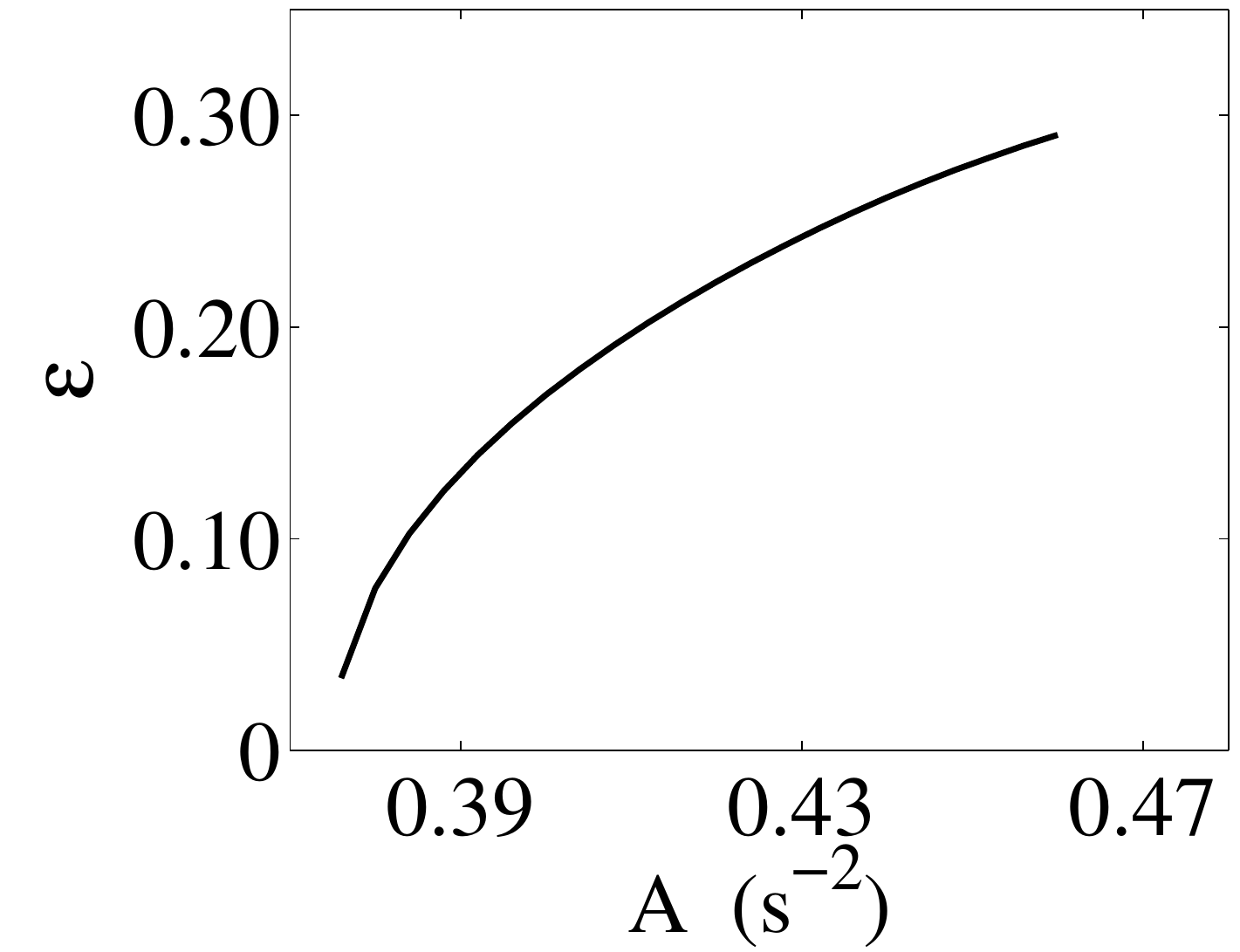}\hspace{2mm}
\includegraphics[height=1.15in]{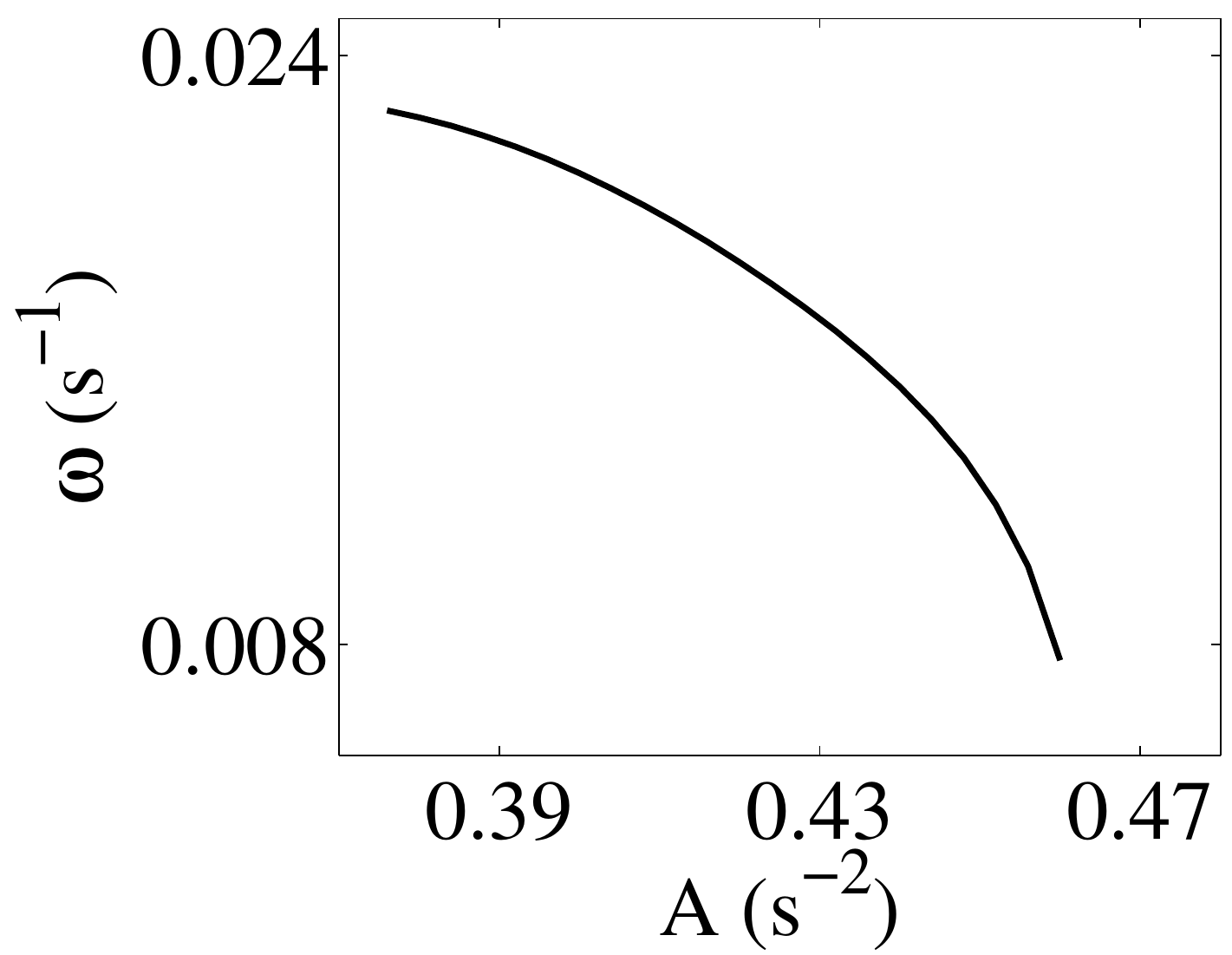}\vspace{2mm}
(b)
\includegraphics[height=1.15in]{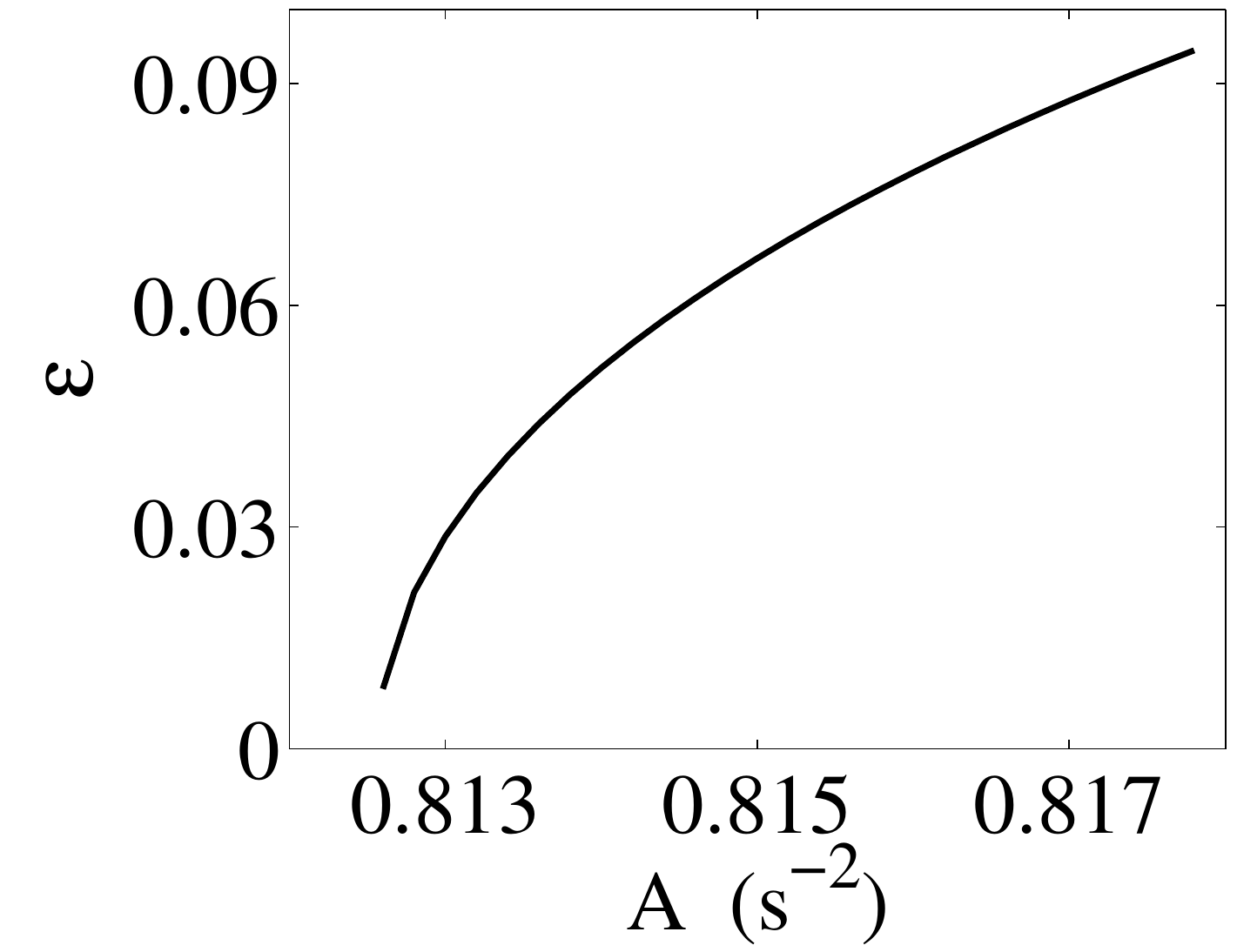}\hspace{2mm}
\includegraphics[height=1.15in]{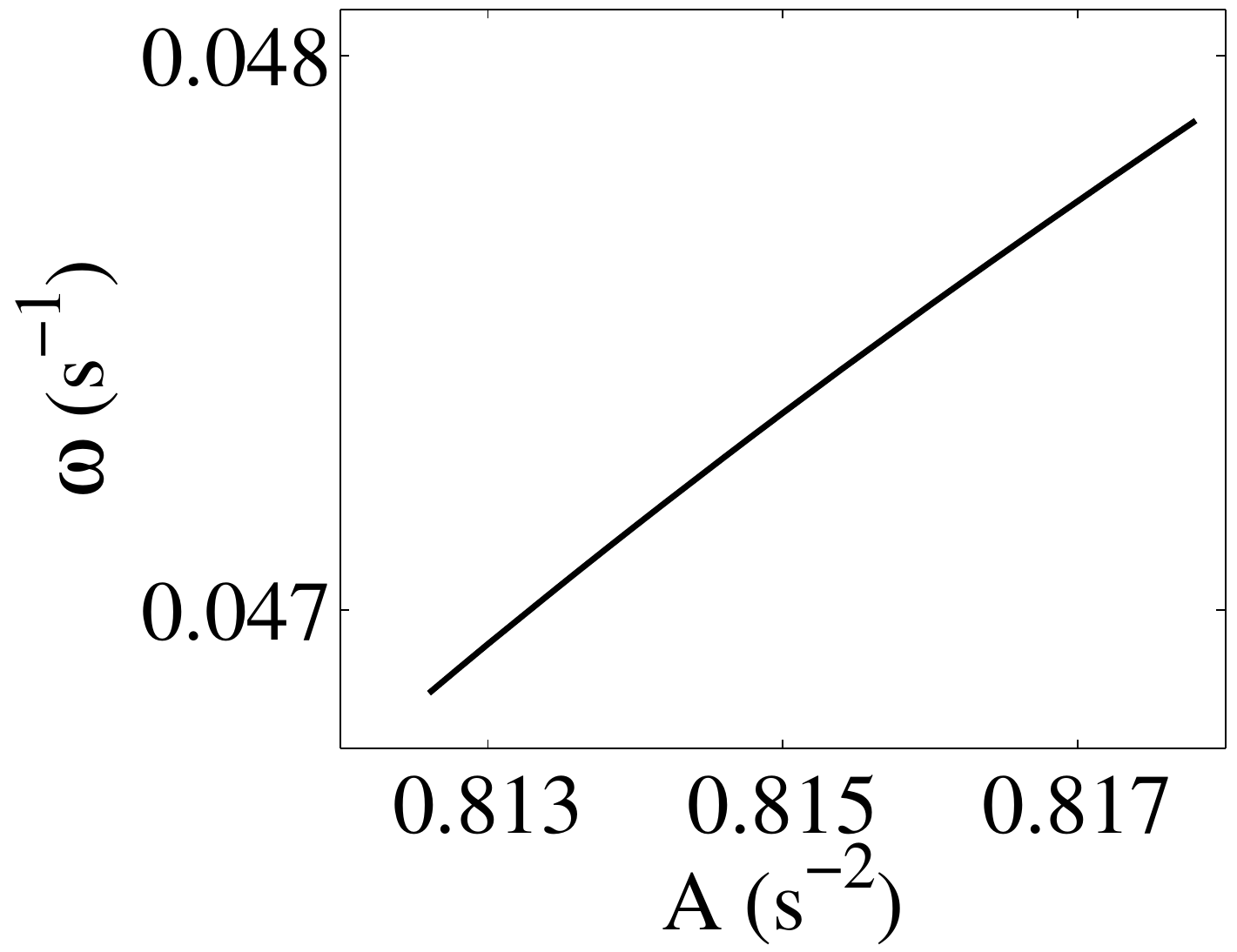}\vspace{2mm}
\hspace{5mm}
(c)
\includegraphics[height=1.15in]{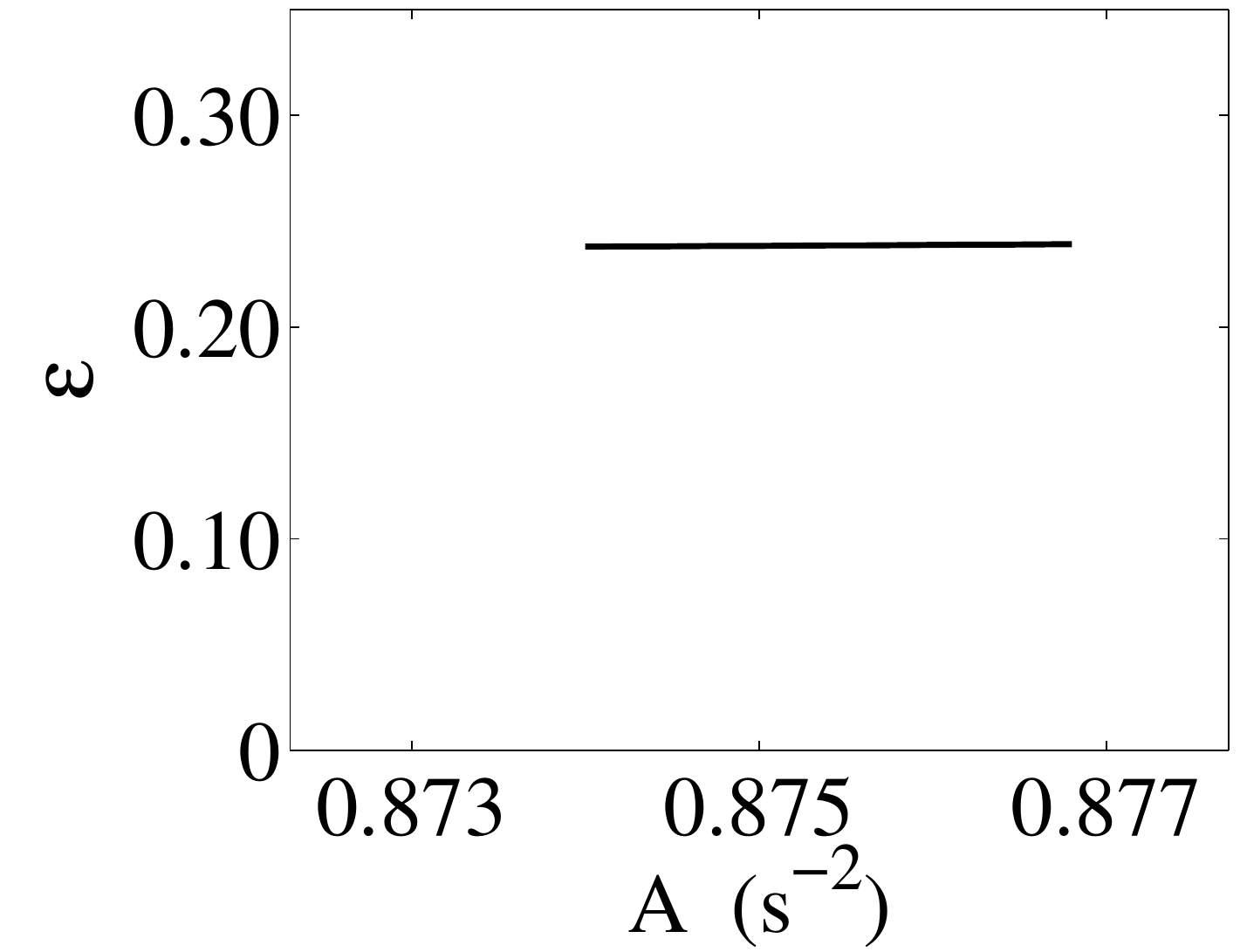}\hspace{2mm}
\includegraphics[height=1.15in]{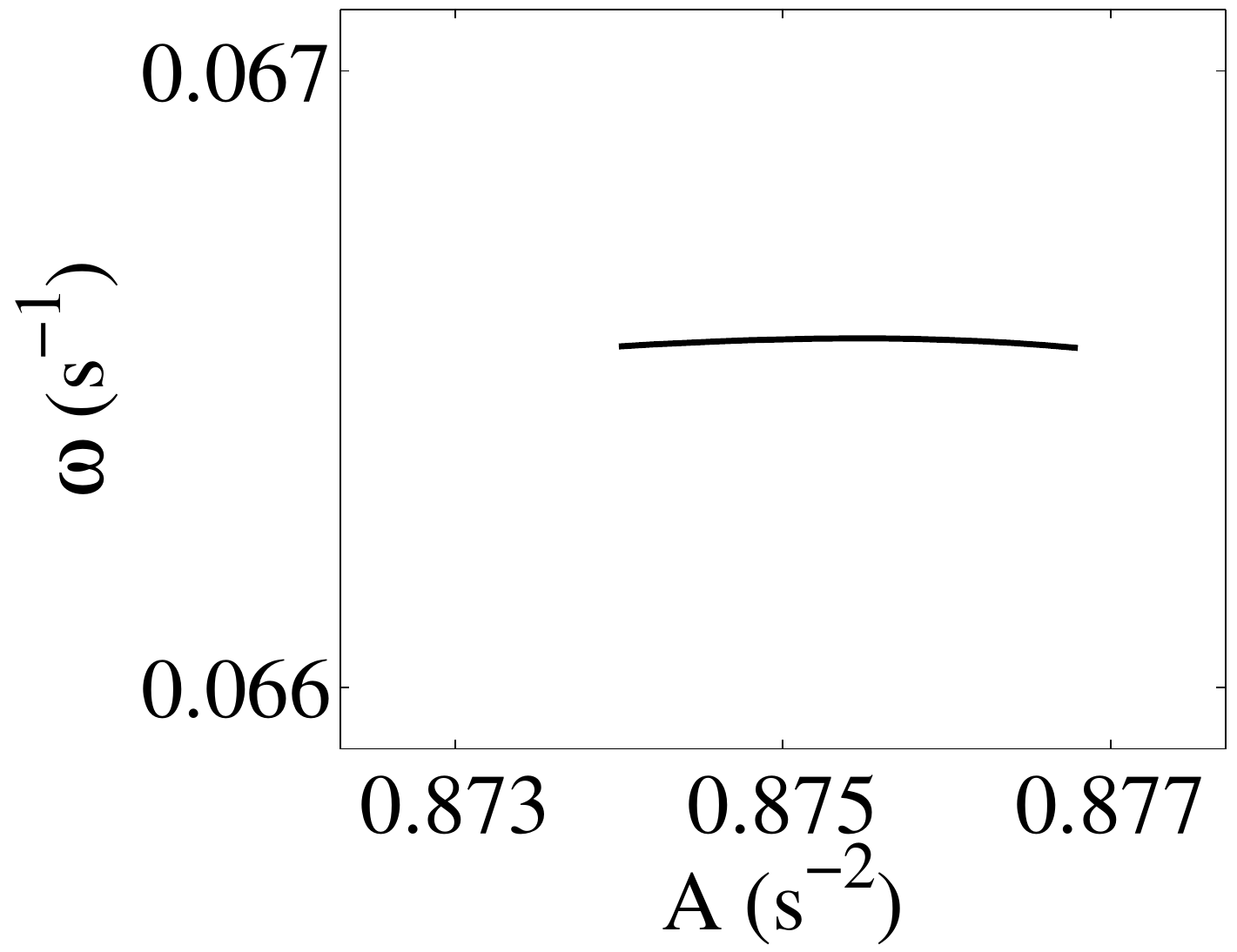}
\caption{\label{P123}The perturbation amplitude $\epsilon$ and frequency $\omega_1=2\pi/T$ of the time-periodic flows $P_1$ (a), $P_2$ (b), and $P_3$ (c). Only the ranges of $A$ are shown where these flows exist and are stable.}
\end{figure}

At this point the first stable, time-periodic solution (denoted $P_1$) appears. Four snapshots of this state at different phases of the oscillation are shown in Fig. \ref{P1}. 
For time-periodic flows it is convenient to represent the stream function as a perturbation about a steady state
\begin{equation}
\Psi(x,y,t)=\Psi_0(x,y)+\epsilon\Psi_1(x,y,t),\label{perturbation}
\end{equation}
where the perturbation $\Psi_1$ has zero time average and $\langle\|\Psi_1\|_2\rangle_t=\|\Psi_0\|_2$ ($\|\ \|_2$ denotes the 2-norm and $\langle\ \rangle_t$ denotes the time average). The strength $\epsilon$ of the time-dependent perturbation as a function of $A$ is shown in Fig. \ref{P123} along with its frequency $\omega_1$.

As expected for a state created via a Hopf bifurcation, the amplitude of oscillation for $P_1$ grows as a square root of the distance to the bifurcation point (see Fig. \ref{P123}(a)). The frequency of oscillations $\omega_1=2\pi/T$ decreases (and the period $T$ increases) monotonically with $A$ until the oscillatory state is destroyed as a result of an infinite-period bifurcation at $A \approx 0.4635$ s$^{-2}$. 

\begin{figure}
(a)\includegraphics[width=3in]{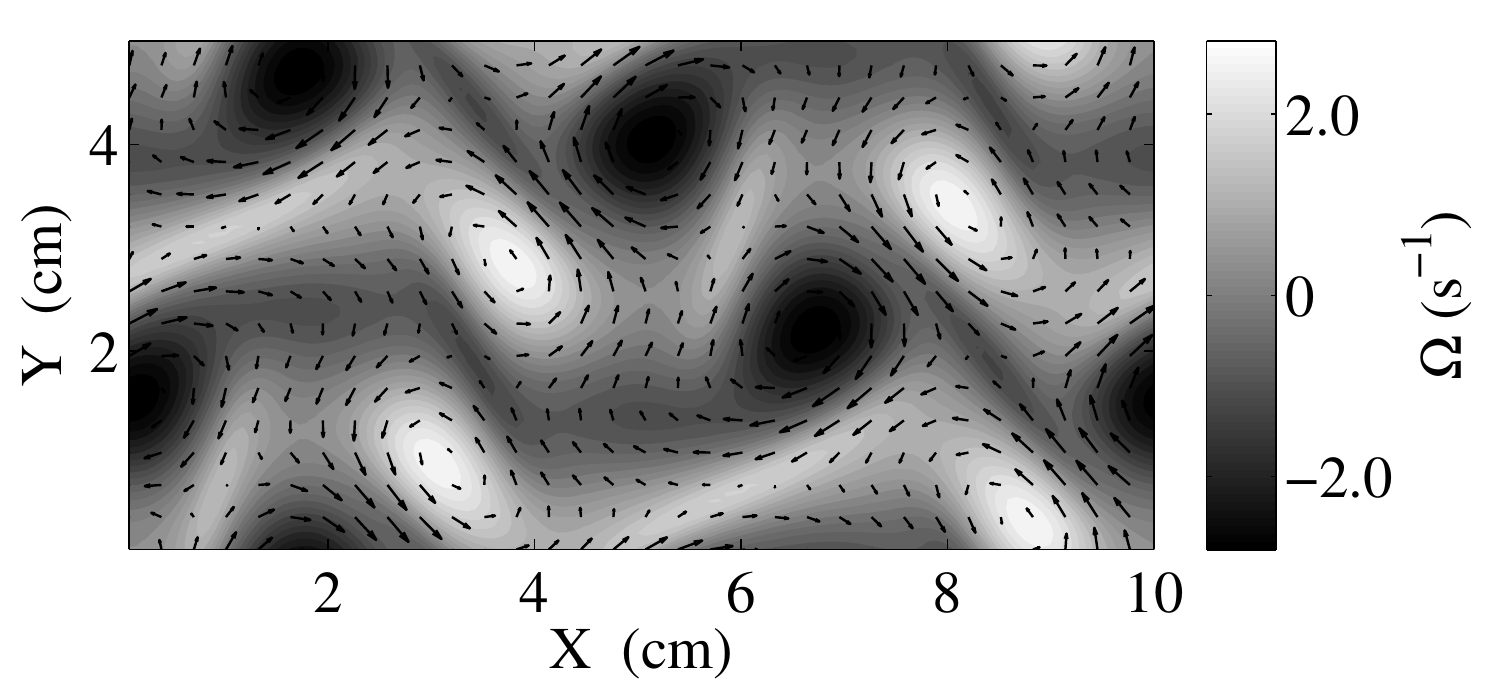}\\
(b)\includegraphics[width=3in]{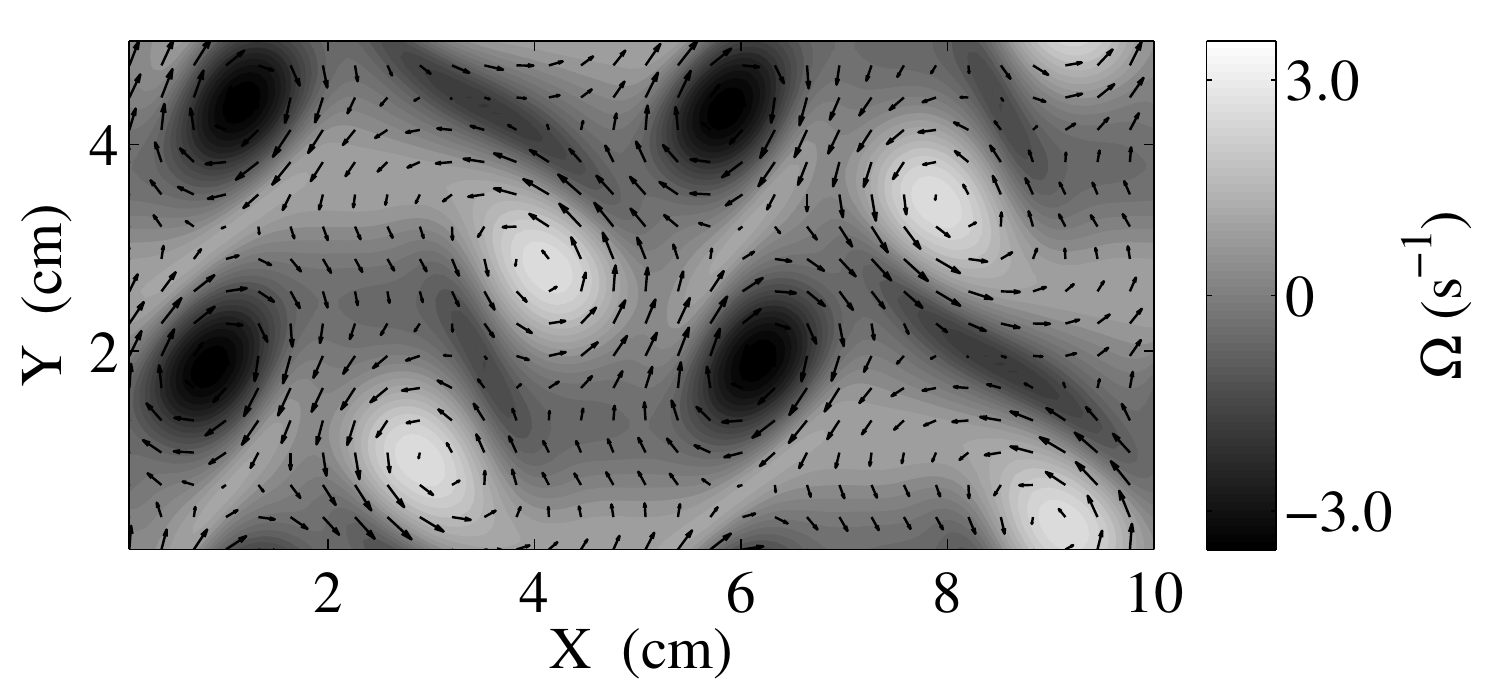}
\caption{Stable steady flow $N$ (a) and unstable steady flow $S$ (b) at $A=0.750$ s$^{-2}$.}
\label{NS}
\end{figure}

At this point two steady solutions are created, a stable node $N$ and a saddle $S$ (shown in Figs. \ref{NS}(a) and (b), respectively). The corresponding flows are quite similar (a disordered array of four clockwise and four counterclockwise vortices) and possess a relatively low symmetry: just like $P_1$, they are symmetric with respect to a shift $(x,y)\to(x+L_x/2,y+L_y/2)$.

\begin{figure}[b]
\includegraphics[width=\columnwidth]{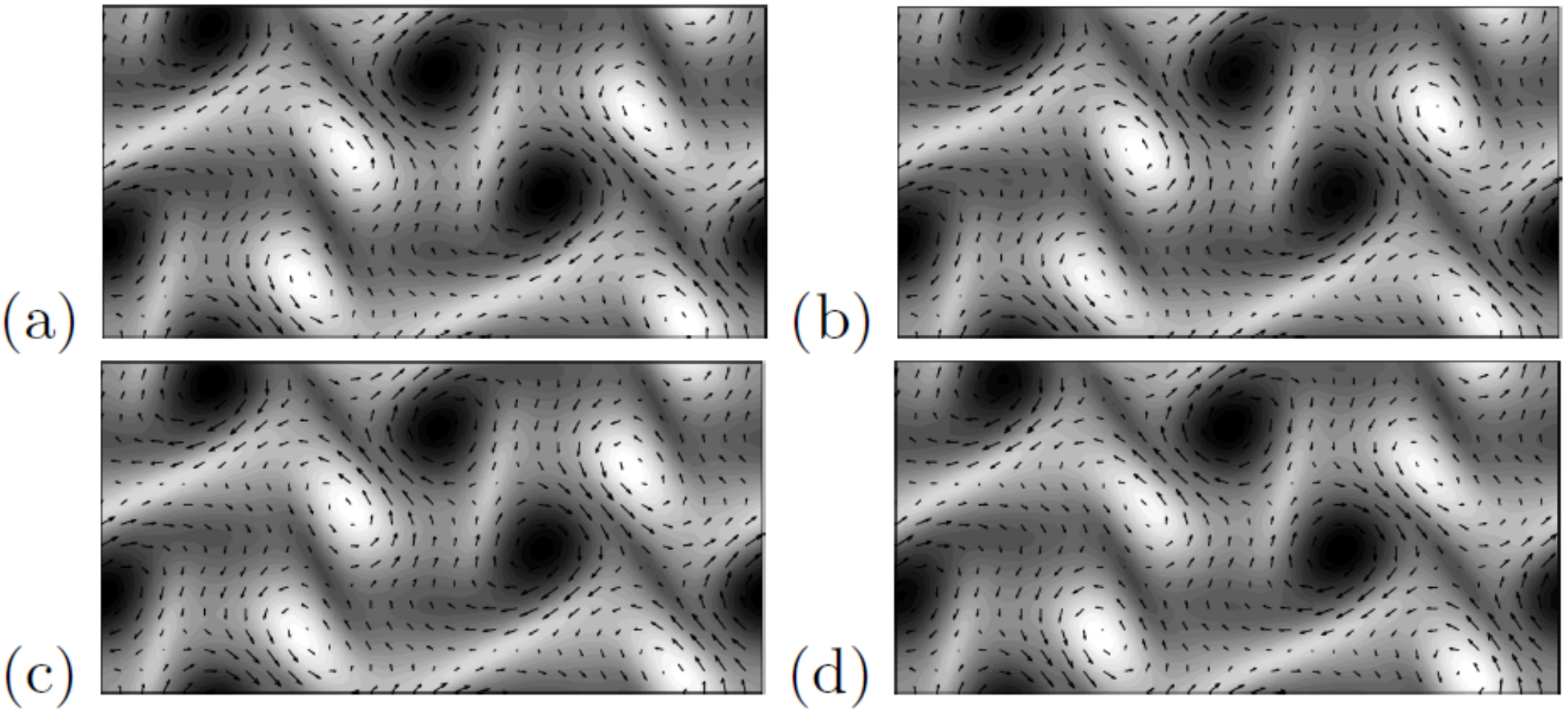}
\caption{Time-periodic flow $P_2$ at $A=0.817$ s$^{-2}$ and (a) t=0, (b) t=T/4, (c) t=T/2, (d) t=3T/4 with $T=131.76$ s. The same color bar as in Fig. \ref{NS}(a) is used here.}
\label{P2}
\end{figure}

The numerical solution of (\ref{vorticity}) follows the stable branch $N$ as $A$ increases further until the corresponding steady flow again develops an oscillatory instability (also a supercritical Hopf) at $A \approx 0.8125$ s$^{-2}$, giving rise to another time-periodic flow $P_2$, shown in Fig. \ref{P2}. The amplitude and frequency of this flow are shown in Fig. \ref{P123}(b). This state is stable in a fairly narrow range of $A$ and, at $A \approx 0.8180$ s$^{-2}$, $P_2$ undergoes a secondary supercritical Hopf bifurcation giving rise to a quasi-periodic flow (denoted $QP$) which, after another Hopf bifurcation, transitions to aperiodic flow around $A \approx 0.865$ s$^{-2}$.

\begin{figure}
\includegraphics[width=\columnwidth]{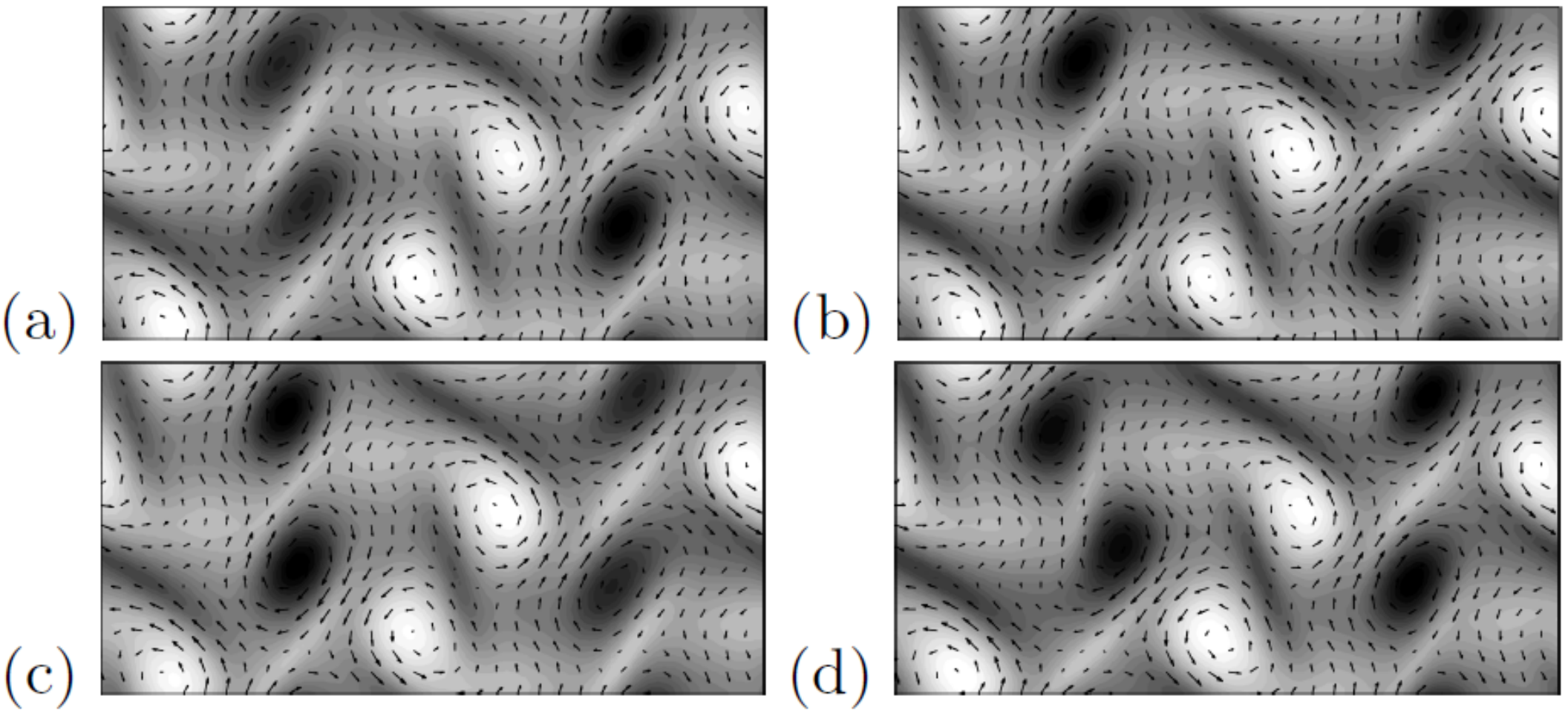}
\caption{Time-periodic flow $P_3$ at $A=0.875$ s$^{-2}$ and (a) t=0, (b) t=T/4, (c) t=T/2, (d) t=3T/4 with $T=94.39$ s. The same color bar as in Fig. \ref{NS}(b) is used here.}
\label{P3}
\end{figure}

At $A \approx 0.8740$ s$^{-2}$, a third stable, time-periodic state $P_3$, shown in Fig. \ref{P3}, is created via a subcritical Hopf bifurcation. The corresponding flow does not respect any of the symmetries of the system and is only stable for a very narrow range of $A$ before it undergoes a subcritical pitchfork bifurcation at $A \approx 0.8768$ s$^{-2}$. Its amplitude and frequency are effectively constant throughout its range of stability as Fig. \ref{P123}(c) illustrates.

Increasing $A$ further, we find another narrow aperiodic window before the flow returns to quasi-periodic behavior at about $A \approx 0.885$ s$^{-2}$. Finally, the flow once again becomes aperiodic at $A \approx 0.980$ s$^{-2}$. The temporally aperiodic (or chaotic) flows we find are weakly turbulent.

\begin{figure}
\includegraphics[width=3in]{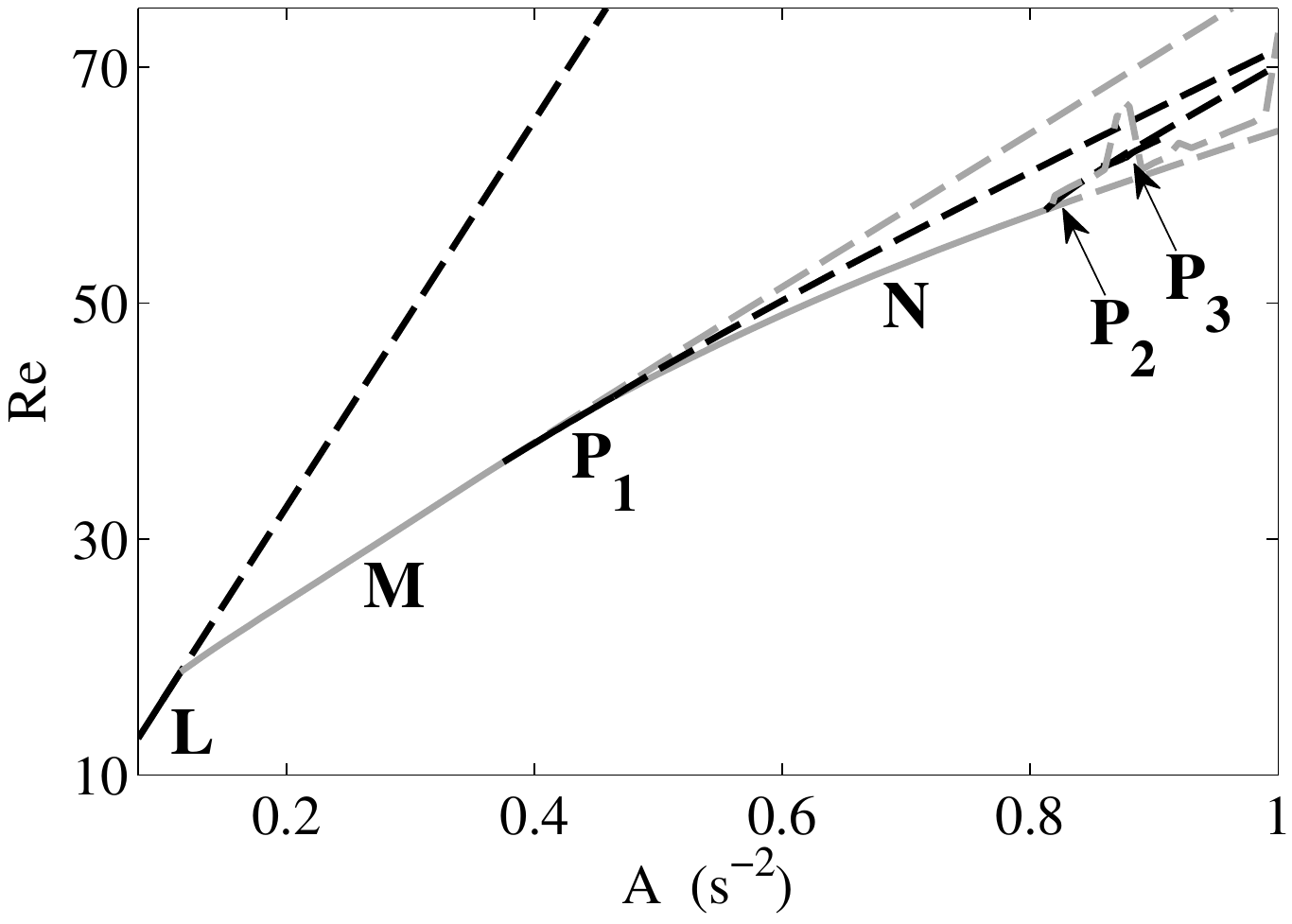}
\caption{Relationship between the Reynolds number and the forcing strength $A$. Again solid lines denote stable states and dashed lines denote unstable ones.}
\label{RevsA}
\end{figure}

We conclude this section by a discussion of the Reynolds number
\begin{equation}
Re=w \nu^{-1}\|\langle{\bf v}\rangle_t \|_2
\end{equation}
characterizing the solutions described above. As Fig. \ref{RevsA} shows, $Re$ varies linearly with $A$ in different flow regimes. The slope is roughly the same for almost all flows, except the laminar flow $L$, for which it is much steeper. Indeed, a quick inspection of the vorticity fields shows that, beyond $L$, the flow is dominated by structures oriented at an angle $\theta\approx 45$ degrees to the $x$ direction, so that the slope can be estimated as $Re/A\sim (k/\sin\theta)^{-4} \nu^{-2}\approx 47.5$ s$^2$. For the laminar flow we find instead $Re/A\sim k^{-4} \nu^{-2}\approx 190$ s$^2$. Both estimates are in reasonable agreement with the numerical data presented in Fig. \ref{RevsA}.

\section{Mixing Properties}
\label{s:mixing}

\subsection{Numerical Results}

In order to quantify the transport properties of the flow, it would be convenient to use two different metrics: (i) the relative size (in this case area) of the mixed region and (ii) the rate of mixing. Both metrics are most easily evaluated by following the evolution of an initially well-localized array of passive tracers. Before continuing with the detailed discussion of mixing dynamics, we should point out that, while the laminar flow $L$ is expected to be the worst mixer and the aperiodic (turbulent) flow to be the best, the complicated sequence of transitional states observed as $A$ is increased implies that we should not expect a monotonic increase for either metric. While one would expect both metrics to mirror the spatial and temporal complexity of the flow, we find that this correlation is far from perfect.

As discussed in the introduction, the dynamics of passive tracers (\ref{tracers}) is formally Hamiltonian, with $\Psi$ being the Hamiltonian. Time-independent, one-degree-of-freedom Hamiltonian systems are always integrable and thus exhibit regular motion. The tracers follow closed stream lines on which $\Psi$ is exactly conserved, hence, the initial tracer distribution eventually stretches along the stream line passing through its center, but never broadens. However, the introduction of time-dependence is expected to split the flow domain into regions of chaotic and regular dynamics. The relation between mixing and chaotic stream lines establishes a direct analogy between transport in one-degree-of-freedom Hamiltonian systems and mixing in 2D area-preserving flows.

In order to quantify the mixing process, for each value of $A$, a set of passive tracers was initially placed in a square region with the side of $0.1$ mm (which corresponds to initial area fraction $f(0)=2\times 10^{-6}$). Since the greatest degree of stretching usually occurs along homo- or heteroclinic trajectories, the initial sets were centered on top of one of the saddles of the instantaneous flow field.

Each tracer was then advected by numerically integrating (\ref{tracers}) using a fourth-order, area-preserving, symplectic integrator based on the 2-stage Gauss-Legendre scheme \cite{Hairer2006}. Velocities for each tracer were computed at each time step using a cubic interpolation scheme on the $64\times 128$ grid in real space.

The dispersion of tracers was then used to compute the mixing metrics. The mixed area fraction $f(t)$ was computed by partitioning the flow domain into a set of small boxes and computing the ratio between the number of boxes $m$ containing at least one tracer to the total number of boxes $k$. When the tracers uniformly cover the domain, the area fraction should be unity. However, if there are $k$ boxes with $n$ randomly distributed tracers, the fraction of boxes containing at least one tracer would on average be $p_{n,k}=1-\exp(-n/k)$. Thus, the measured area fraction for each value of $A$ was normalized by $p_{n,k}$
\begin{equation}
f(t)=\frac{m(n,t)}{kp_{n,k}},
\end{equation}
so that a uniformly distributed set of tracers would give an area fraction of one.

\begin{figure}
\includegraphics[width=3in]{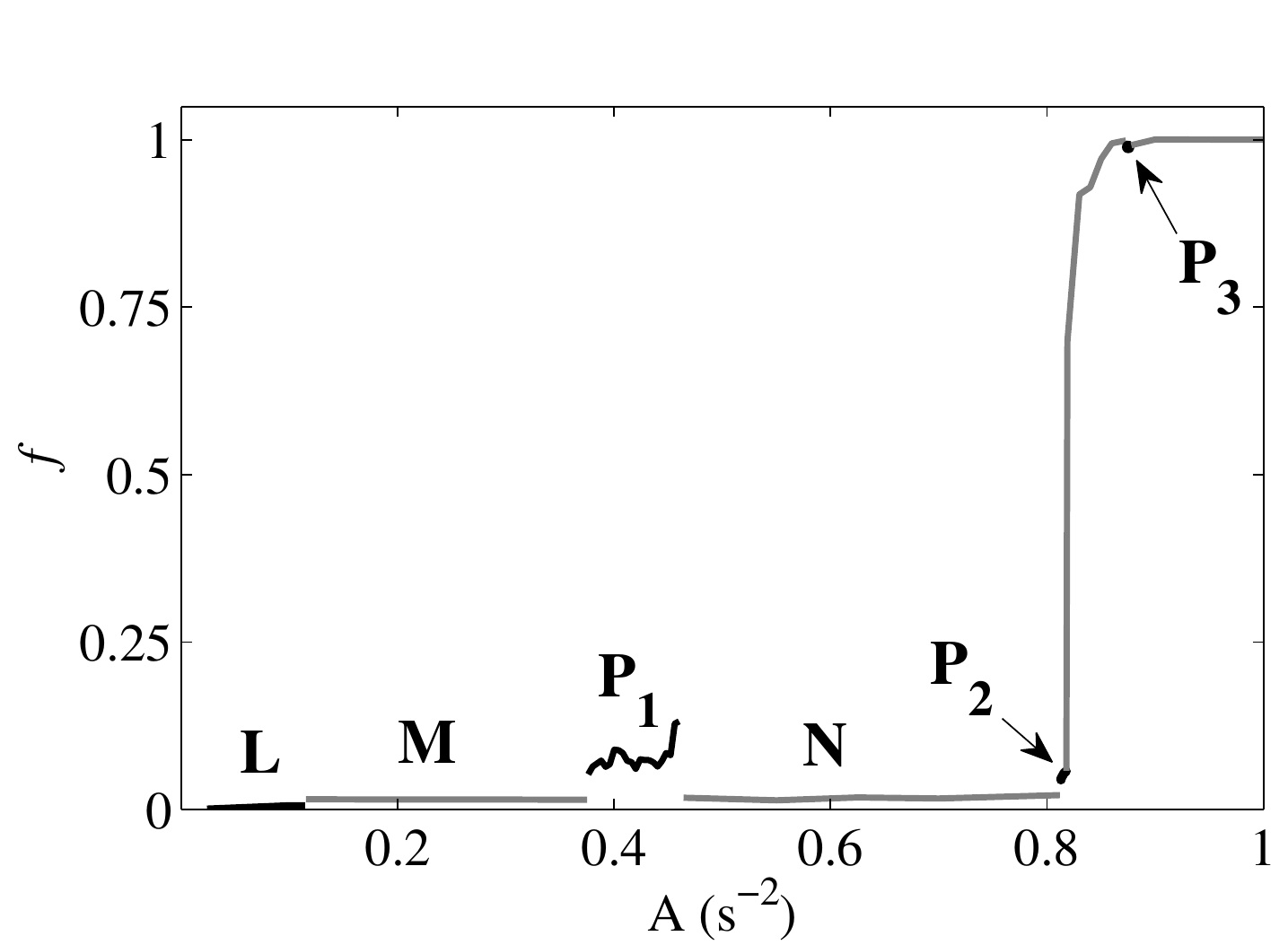}
\caption{The fraction $f$ of the mixed area relative to the total area of the domain at $t=5\times 10^4$ s.}
\label{fvsA}
\end{figure}

Fig. \ref{fvsA} shows the area fraction occupied by the tracers after a rather long time interval of $5\times10^4$ s. In comparison, the period of $P_1$, $P_2$ and $P_3$ is of order $100$ s, while the characteristic time scale of the flow around vortices is below $10$ s. We find that the area fraction remains near zero for all of the time-independent flows ($L$, $M$, and $N$), as it should be for integrable flows. 

For time-dependent flows (\ref{tracers}) formally becomes a three-dimensional dynamical system (augmented by an equation $\dot{t}=1$) which, in general, possesses chaotic solutions (stream lines). Chaotic advection, in principle, should dramatically enhance mixing. However, as Fig. \ref{fvsA} shows, the mixed area fraction for $P_1$ and $P_2$ is only slightly higher than that for the time-independent flows. The time-periodic flow $P_3$, on the other hand, produces nearly perfect mixing, with mixed area fraction comparable to that of aperiodic flows.

\begin{figure}
\centering
(a)\hspace{-8mm}
\includegraphics[width=3.1in]{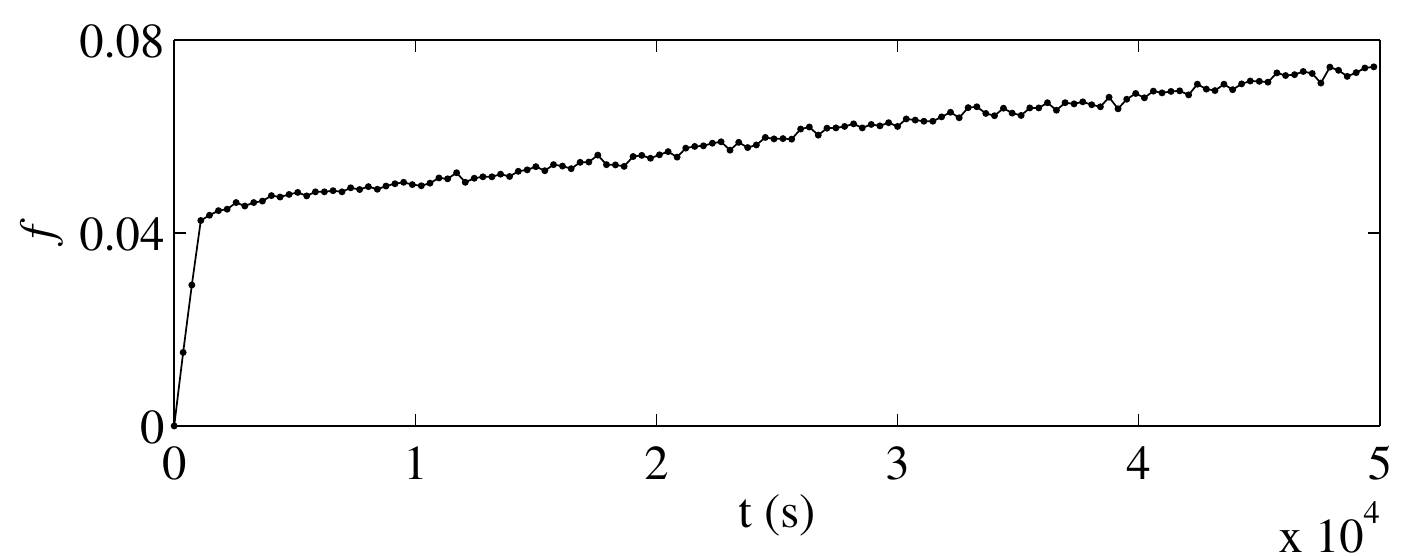}
\\
(b)\hspace{-8mm}
\includegraphics[width=3.1in]{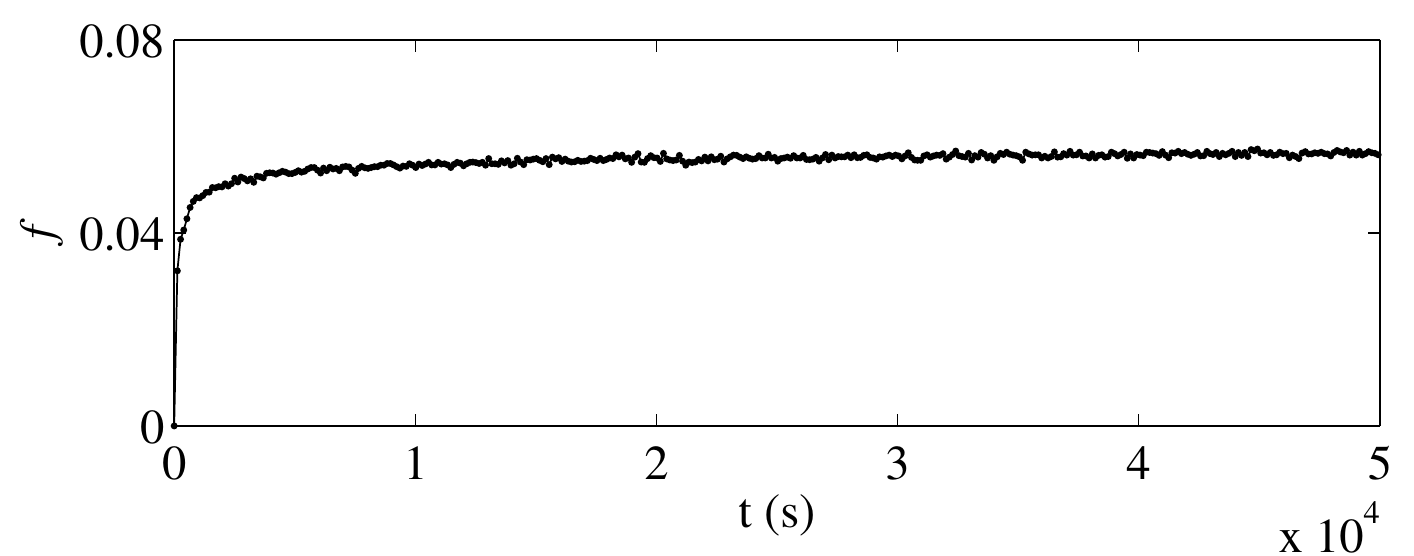}
\\
(c)\hspace{-8mm}
\includegraphics[width=3.0in]{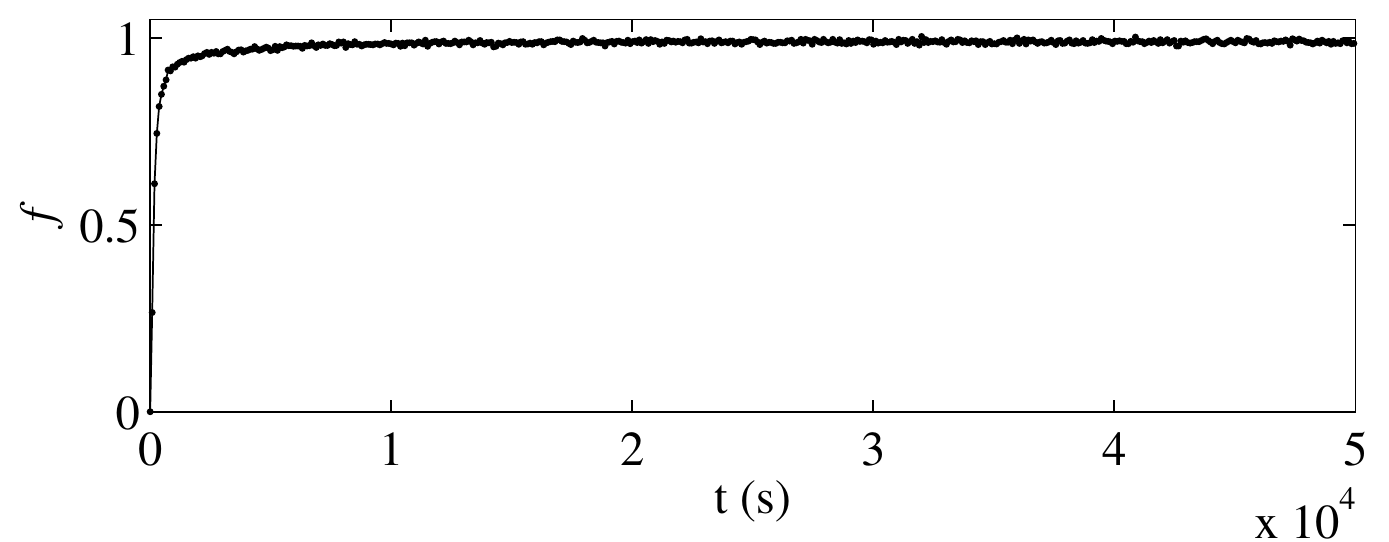}

\caption{\label{fracvst}Temporal dependence of the area fraction for the three time-periodic flows: (a) $P_1$ at $A=0.428$ s$^{-2}$, (b) $P_2$ at $A=0.817$ s$^{-2}$, and (c) $P_3$ at $A=0.875$ s$^{-2}$.}
\end{figure}

\begin{figure*}
\centering{
(a)
\includegraphics[width=3in]{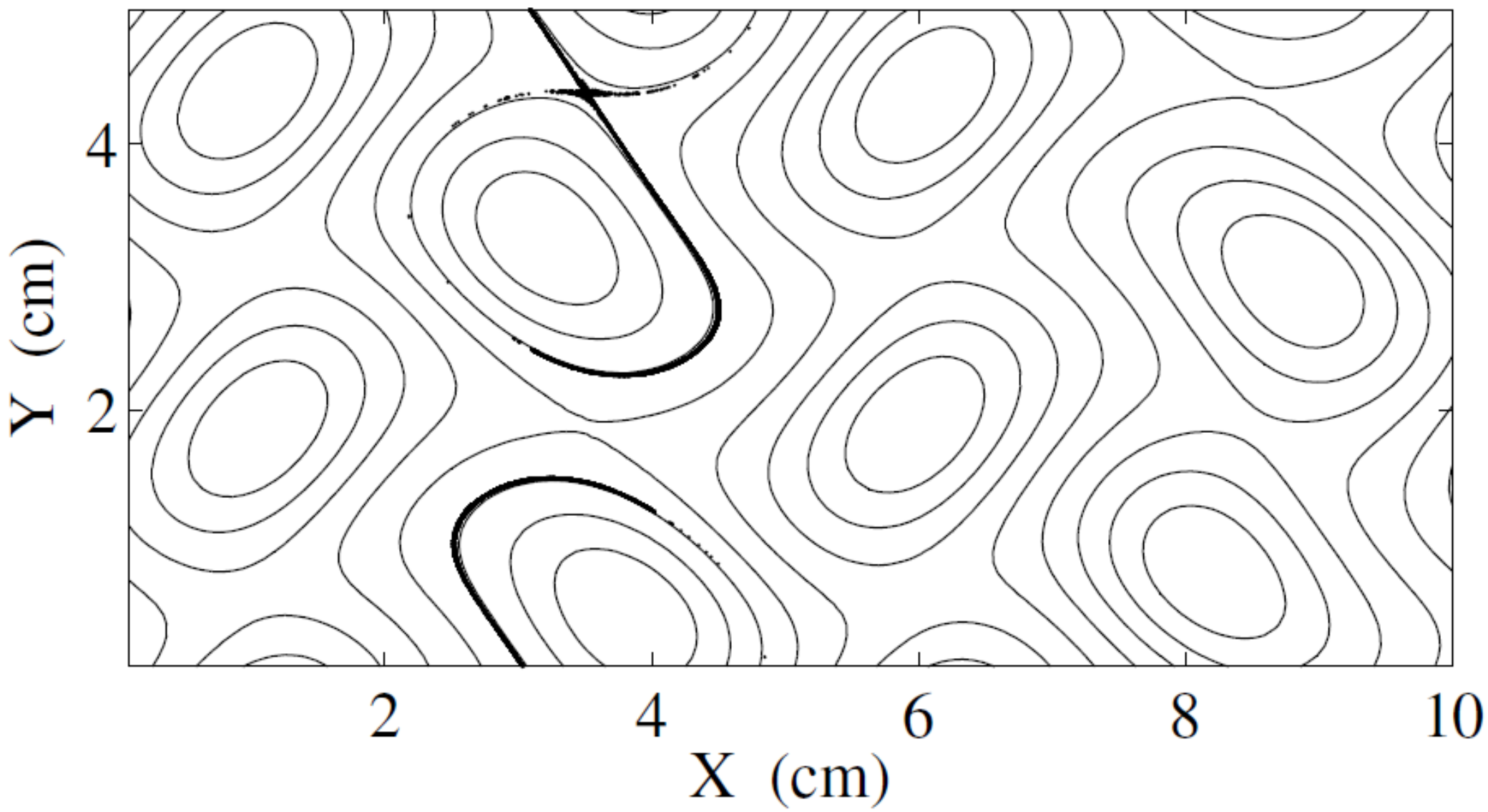}\hspace{5mm}
(b)
\includegraphics[width=3in]{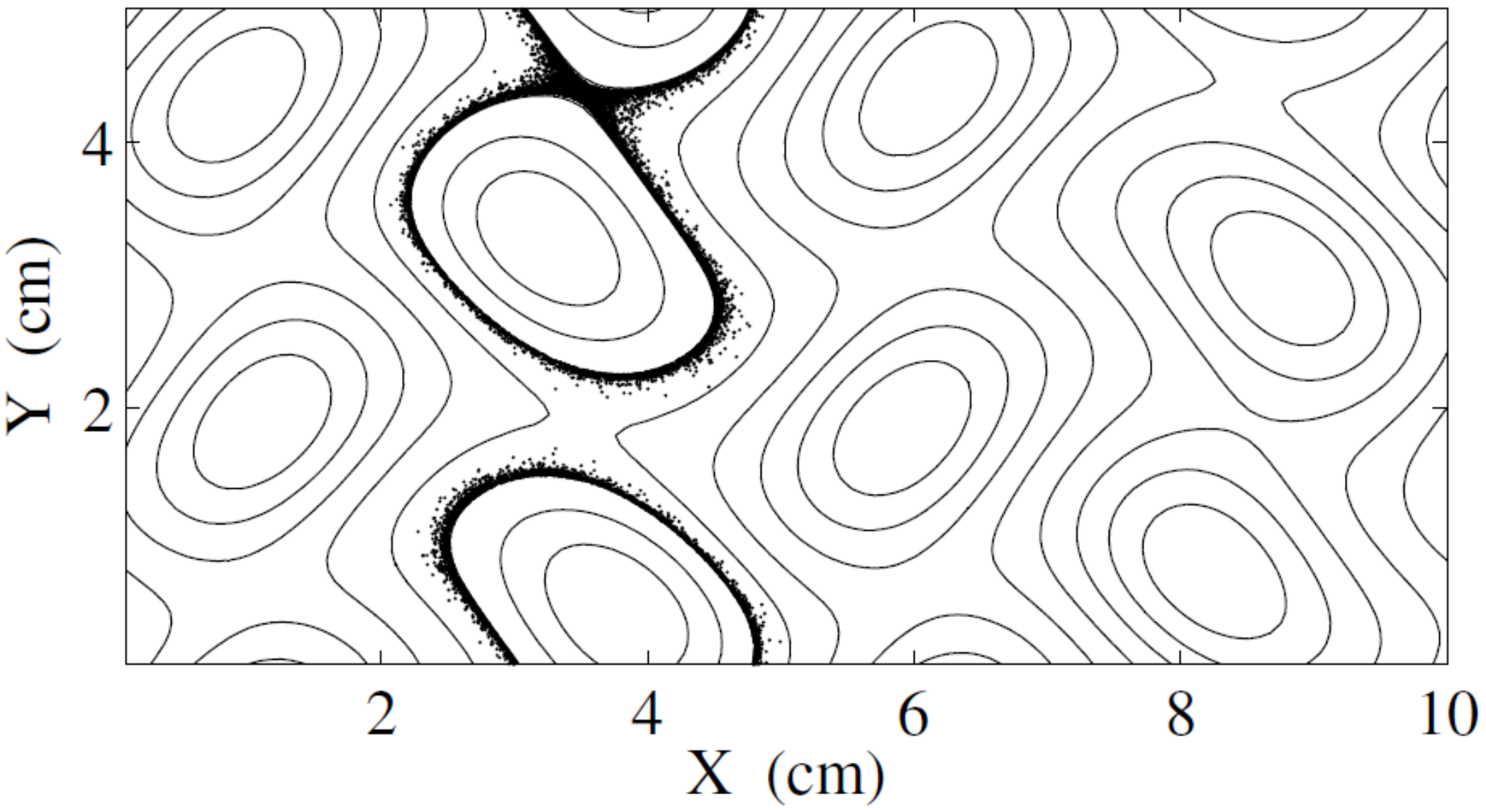}
\\
(c)
\includegraphics[width=3in]{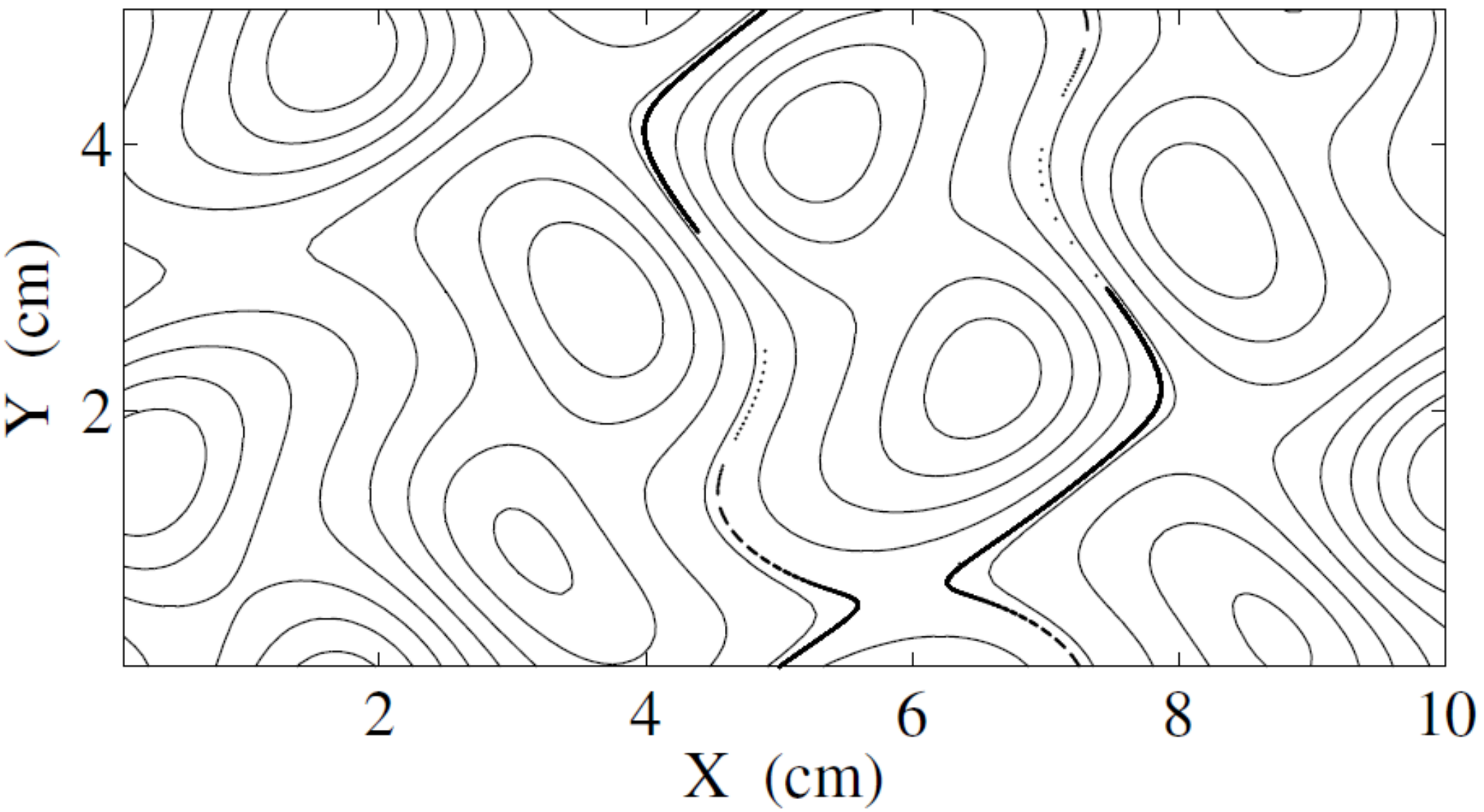}\hspace{5mm}
(d)
\includegraphics[width=3in]{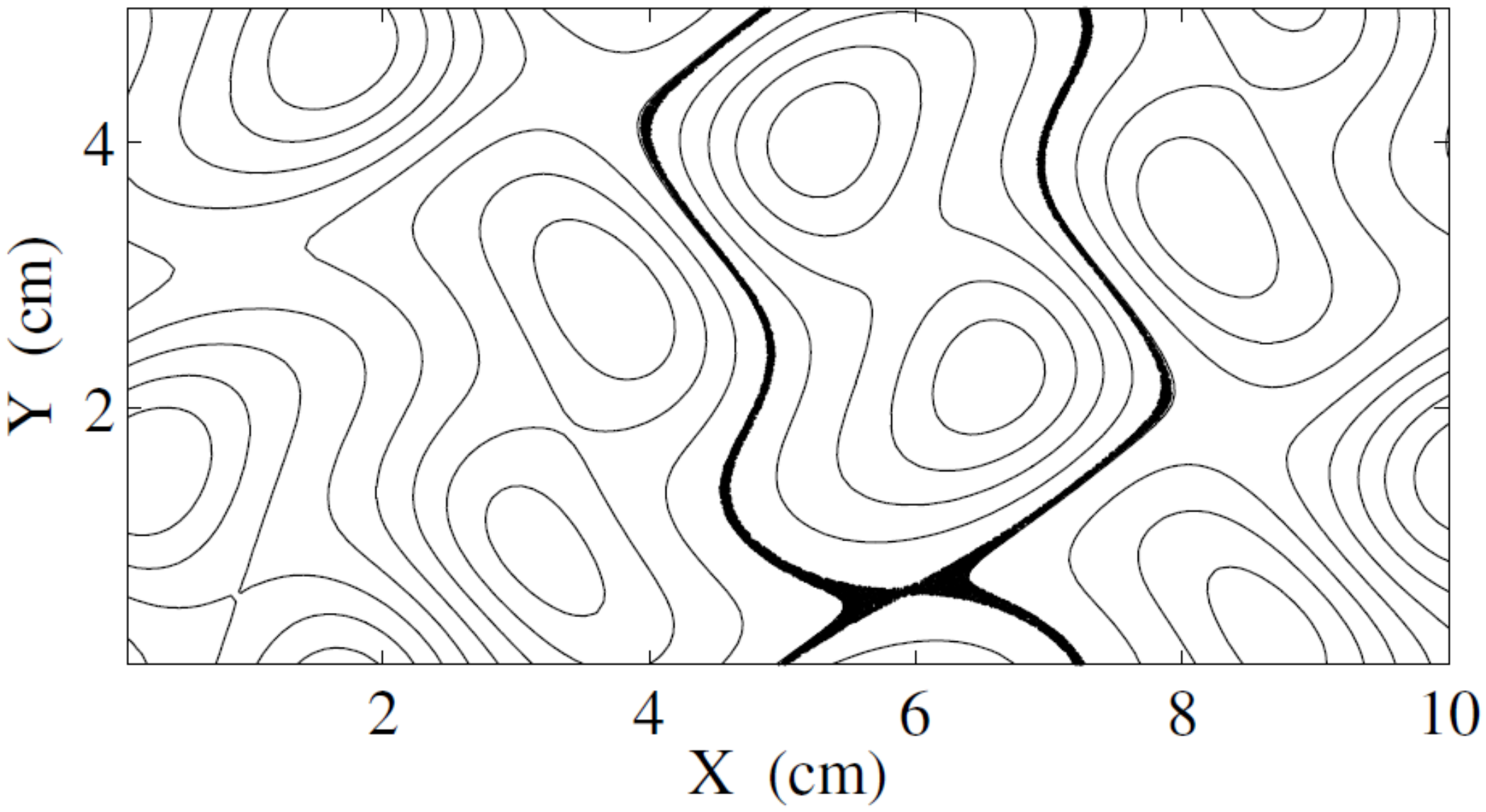}}
\caption{\label{P12-mixing}Mixing by the time periodic flows. The distribution of $6\times10^4$ tracers and the stream lines of the instantaneous flow for $P_1$ at $t=598$ s (a) and $t=5\times10^4$ s (b). The same for $P_2$ at $t=49$ s (c) and $t=5\times10^4$ s (d).}
\end{figure*}

Examining the temporal evolution of the area fraction covered by the tracers shown in Fig. \ref{fracvst}, one can discern two distinct stages for the time periodic flows. Initially there is a very fast increase. For $P_1$ and $P_2$ it corresponds to rapid stretching of the set of tracers along the homoclinic trajectories forming a thin closed band (see Figs. \ref{P12-mixing}(a) and (c)). This is followed by a much slower growth associated with the broadening of this band. However, even after a very long time, the band of tracers remains quite thin and aligned along the stream lines of the instantaneous flow (see Figs. \ref{P12-mixing}(b) and (d)).

\begin{figure*}
\centering{
(a)
\includegraphics[width=3in]{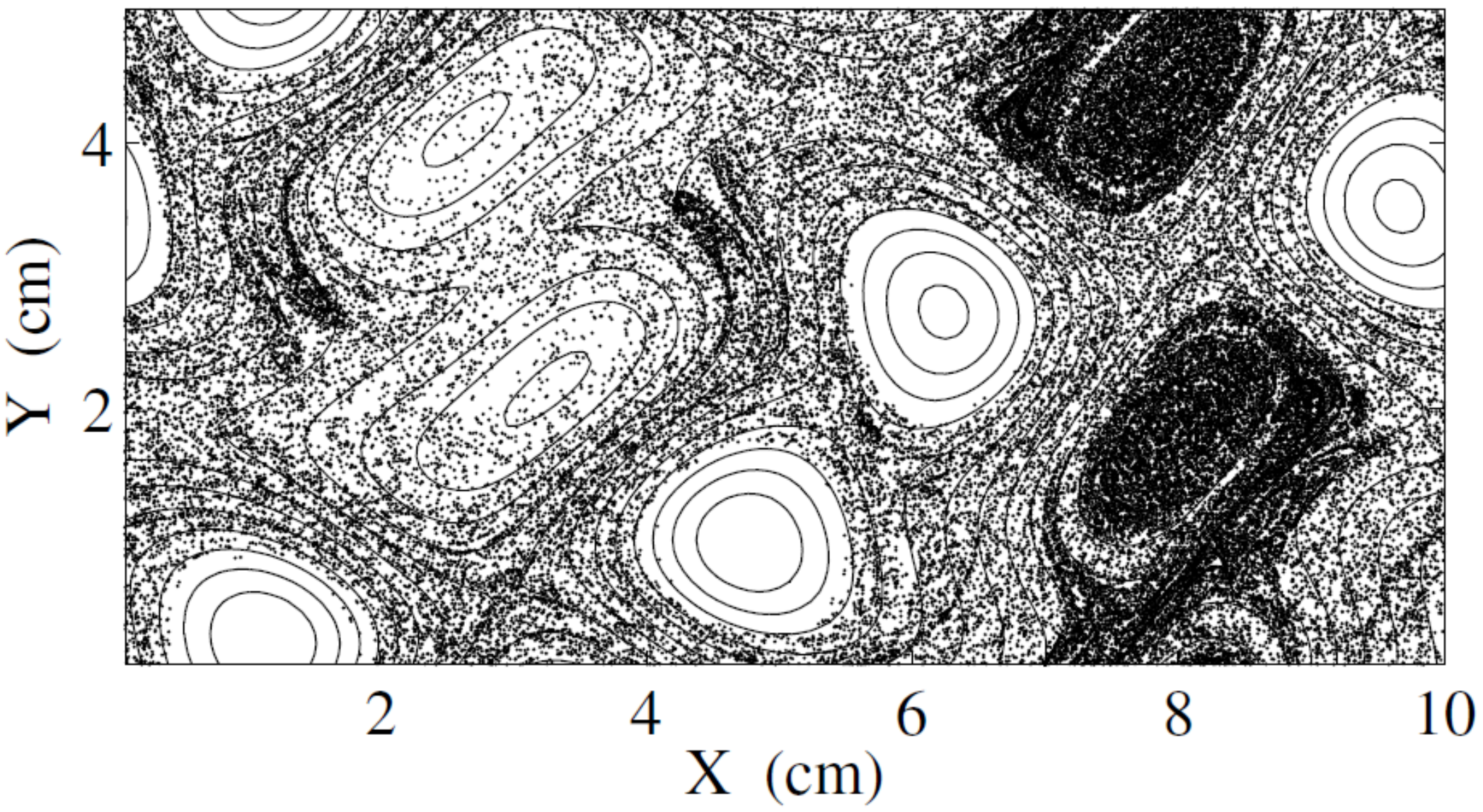}\hspace{5mm}
(b)
\includegraphics[width=3in]{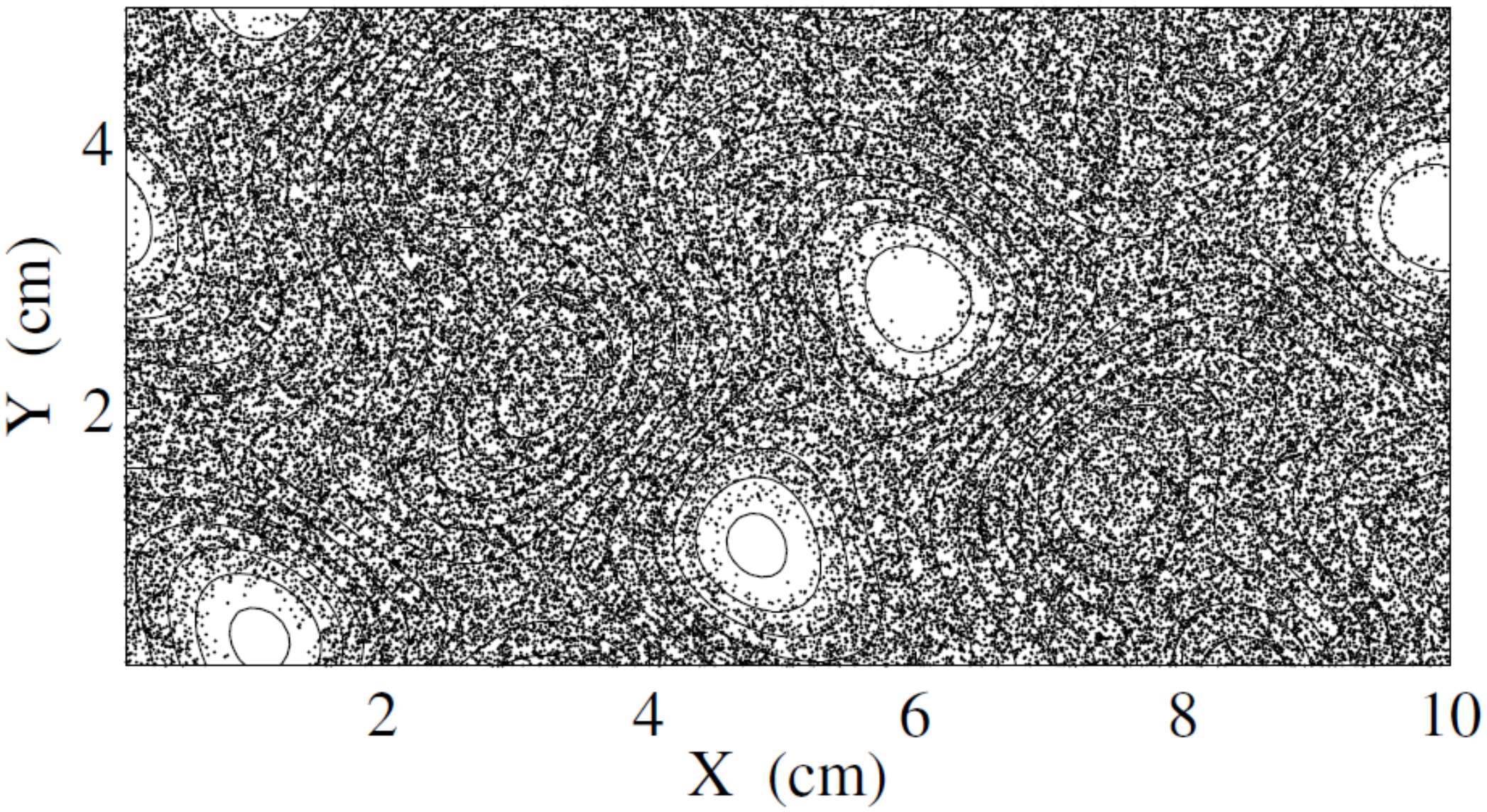}}
\caption{\label{P3-mixing}Mixing by the time-periodic flow $P_3$. The distribution of $6\times10^4$ tracers and the stream lines of the instantaneous flow at $t=317$ s (a) and $t=3500$ s (b).}
\end{figure*}

For $P_3$, on the other hand, the set of tracers undergoes a rapid initial phase of both stretching \textit{and} folding and quickly (within several periods of the flow) covers almost the entire domain (see Fig. \ref{P3-mixing}(a)). A closer look further shows that, for $P_2$ and $P_3$, the tracer distribution reaches an asymptotic state already around $10^3$ s, while for $P_1$ the area fraction is still growing at $t=5\times 10^4$ s. Finally, although the asymptotic distribution of the tracers for $P_3$ is essentially uniform, the tracers never penetrate four small regular islands centered around vortices with positive vorticity, as Fig. \ref{P3-mixing}(b) illustrates. We will return to this fact in Sect. \ref{s:resonance}.

\begin{figure*}
\centering{
(a)
\includegraphics[width=3in]{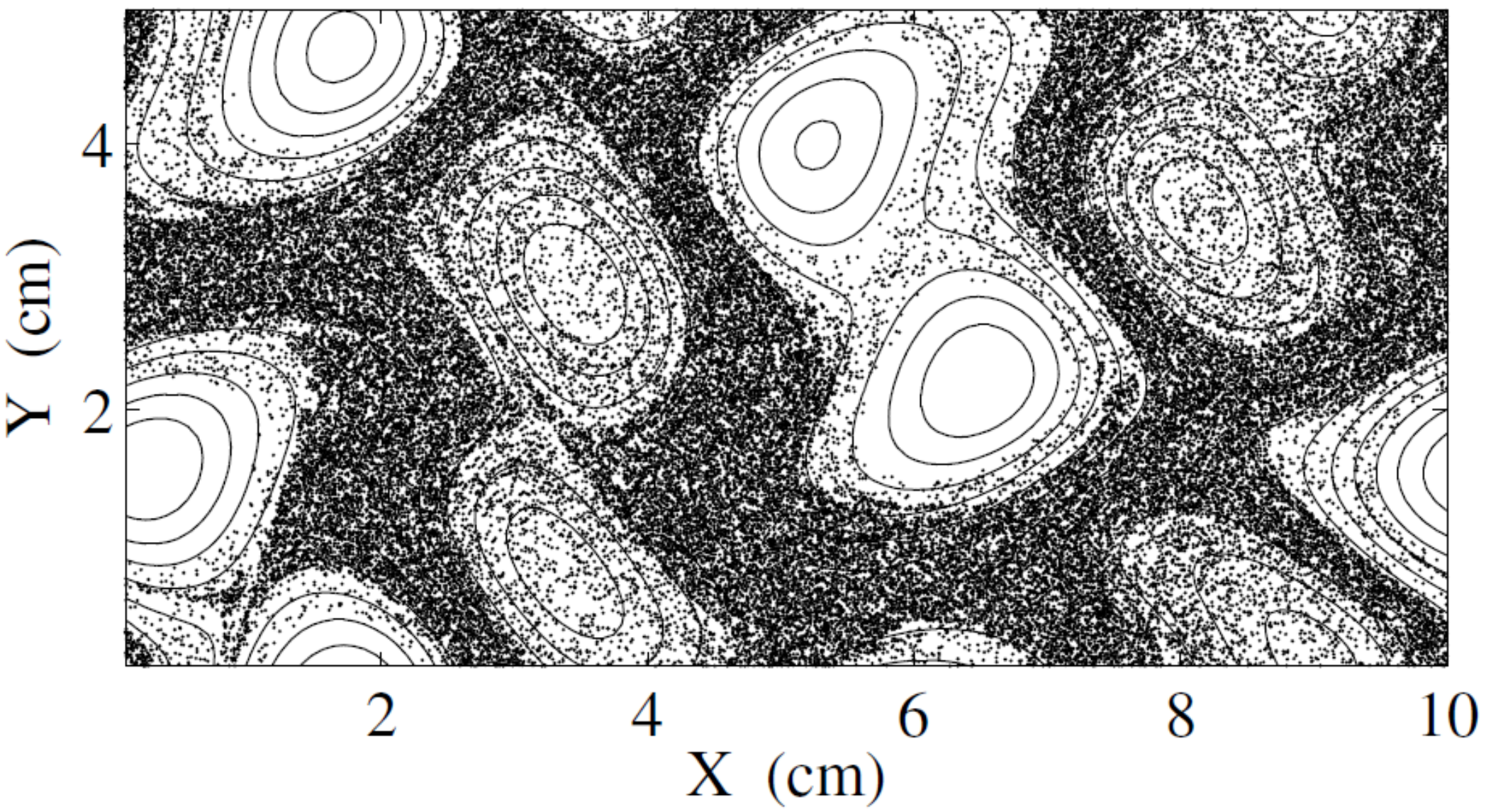}\hspace{5mm}
(b)
\includegraphics[width=3in]{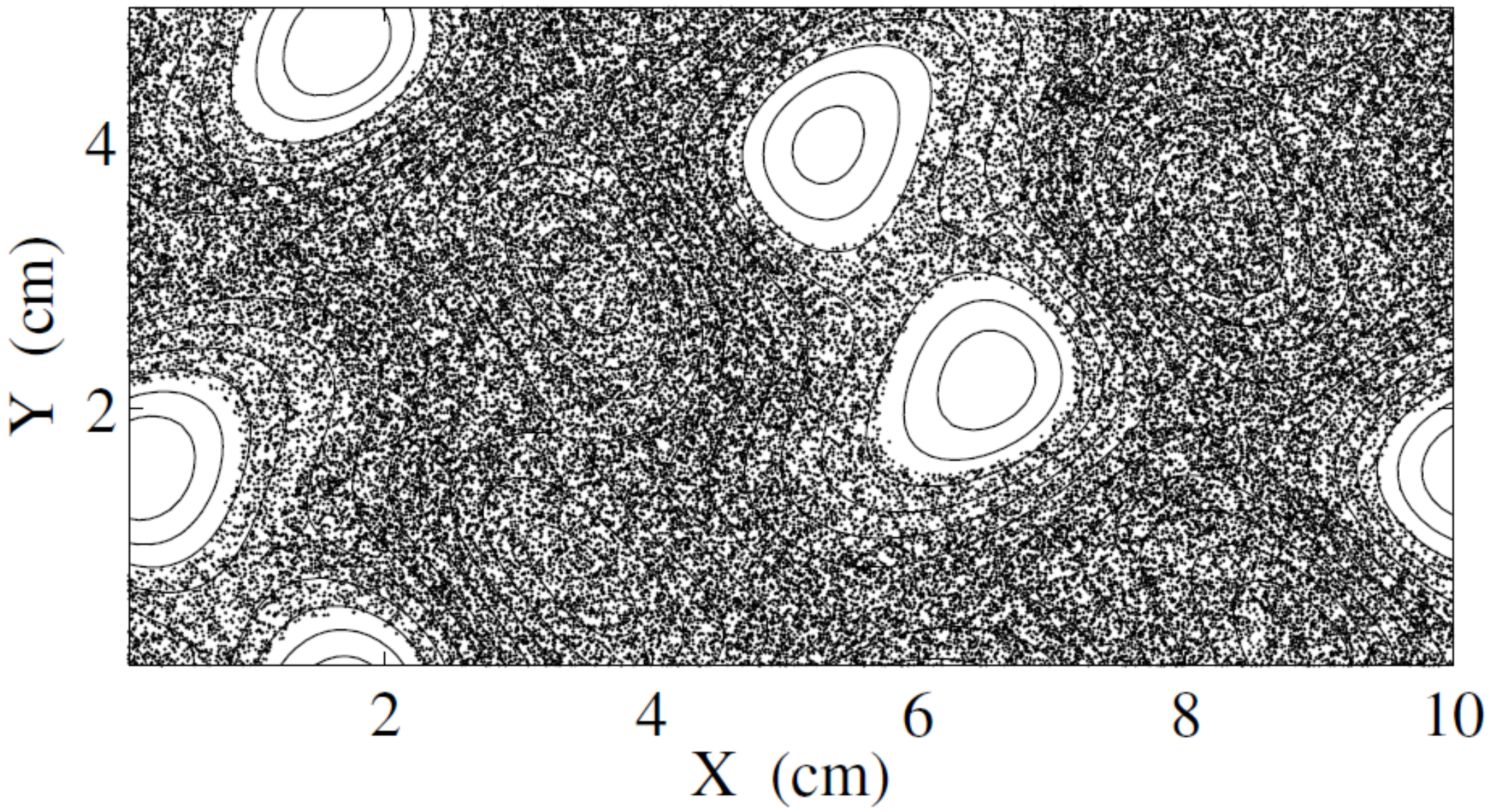}\\
(c)
\includegraphics[width=3in]{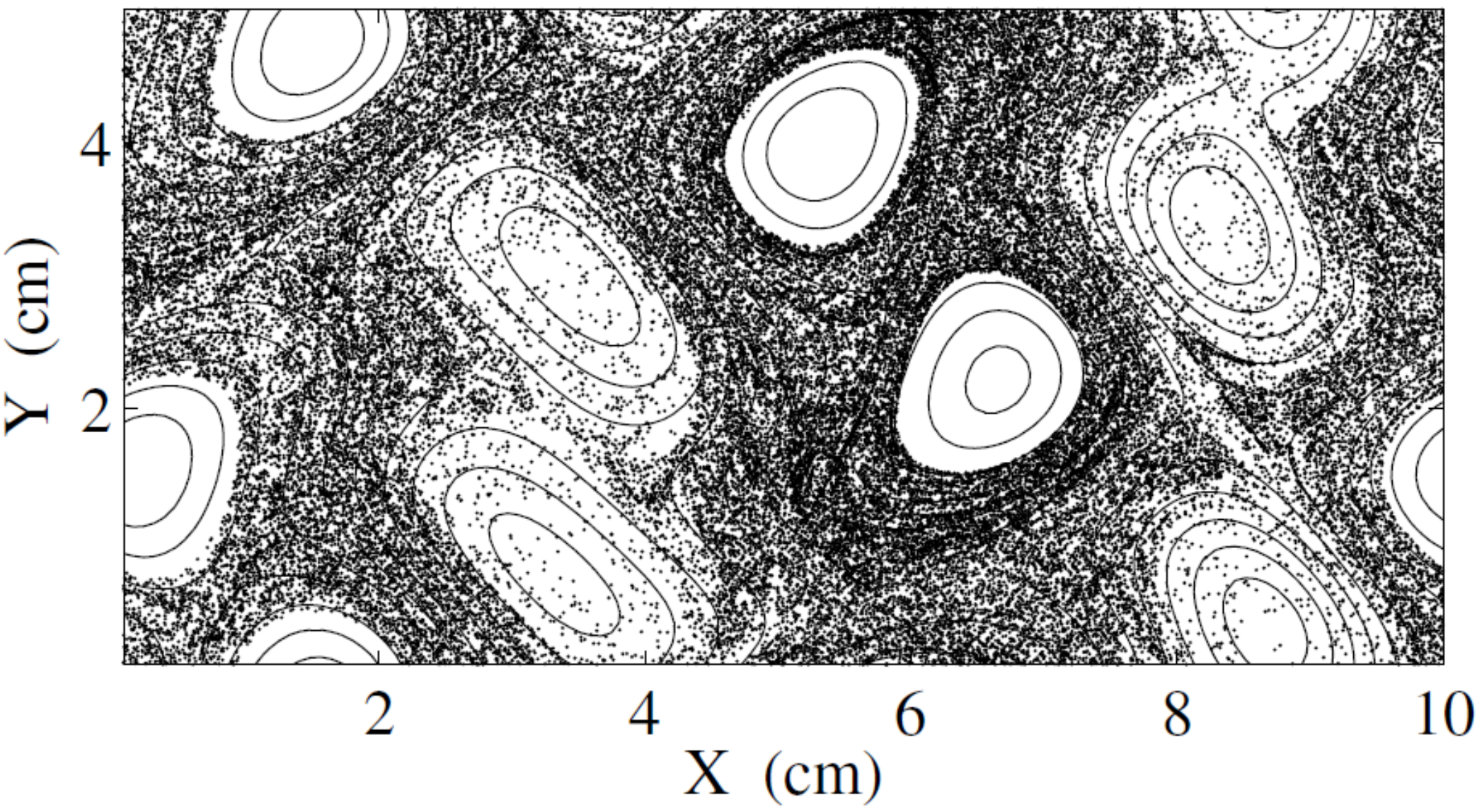}\hspace{5mm}
(d)
\includegraphics[width=3in]{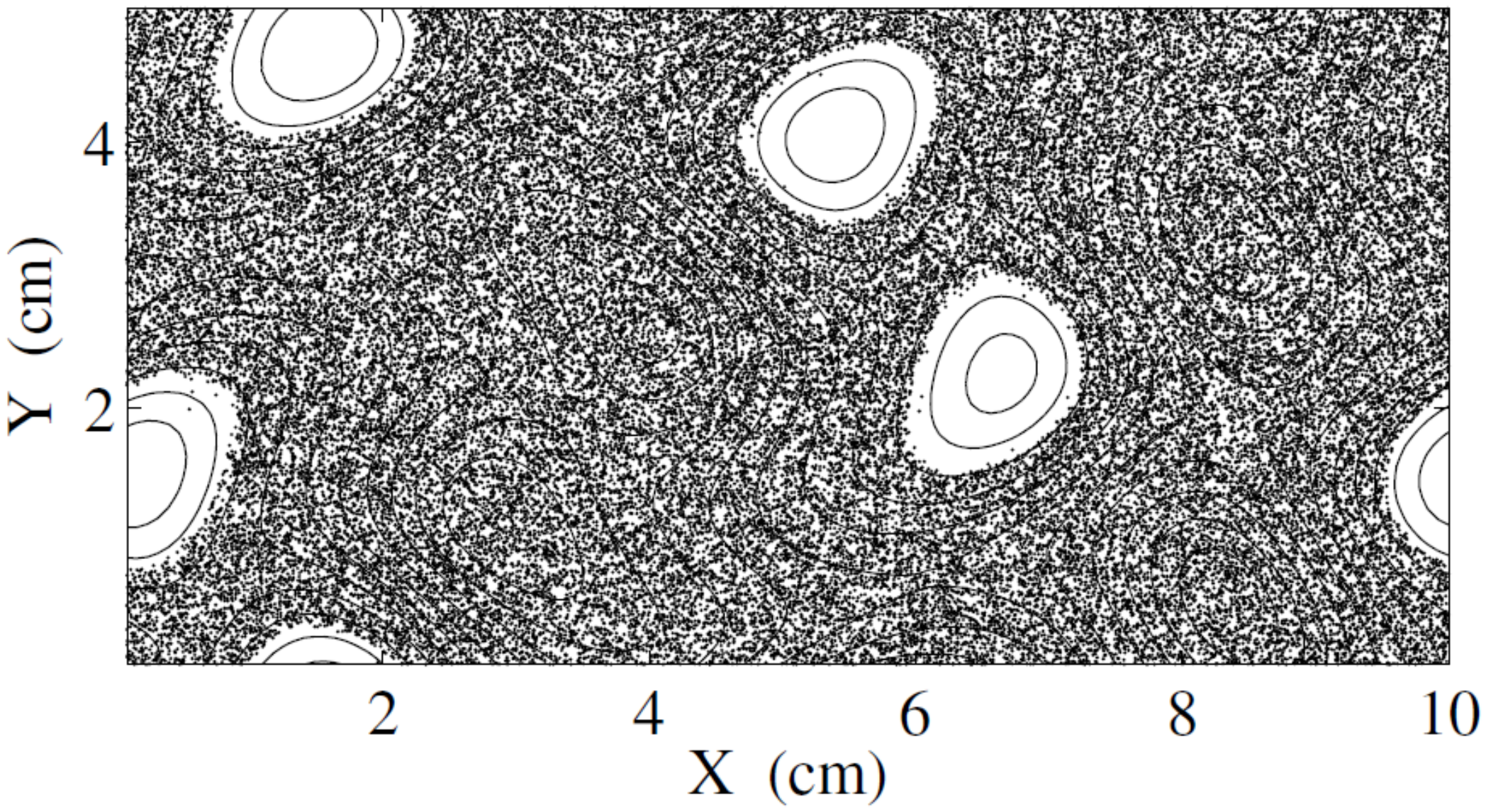}}
\caption{\label{QP-mixing} Mixing by quasi-periodic flow $QP$. The distribution of $6\times10^4$ tracers and stream lines of the instantaneous flow for $A=0.820$ s$^{-2}$ at $t=1038$ s (a) and $t=3500$ s (b). Same for $A=0.846$ s$^{-2}$ at $t=645$ s (c) and $t=3500$ s (d).}
\end{figure*}

\begin{figure*}
\end{figure*}

\begin{figure*}[t]
\centering{
(a)
\includegraphics[width=3in]{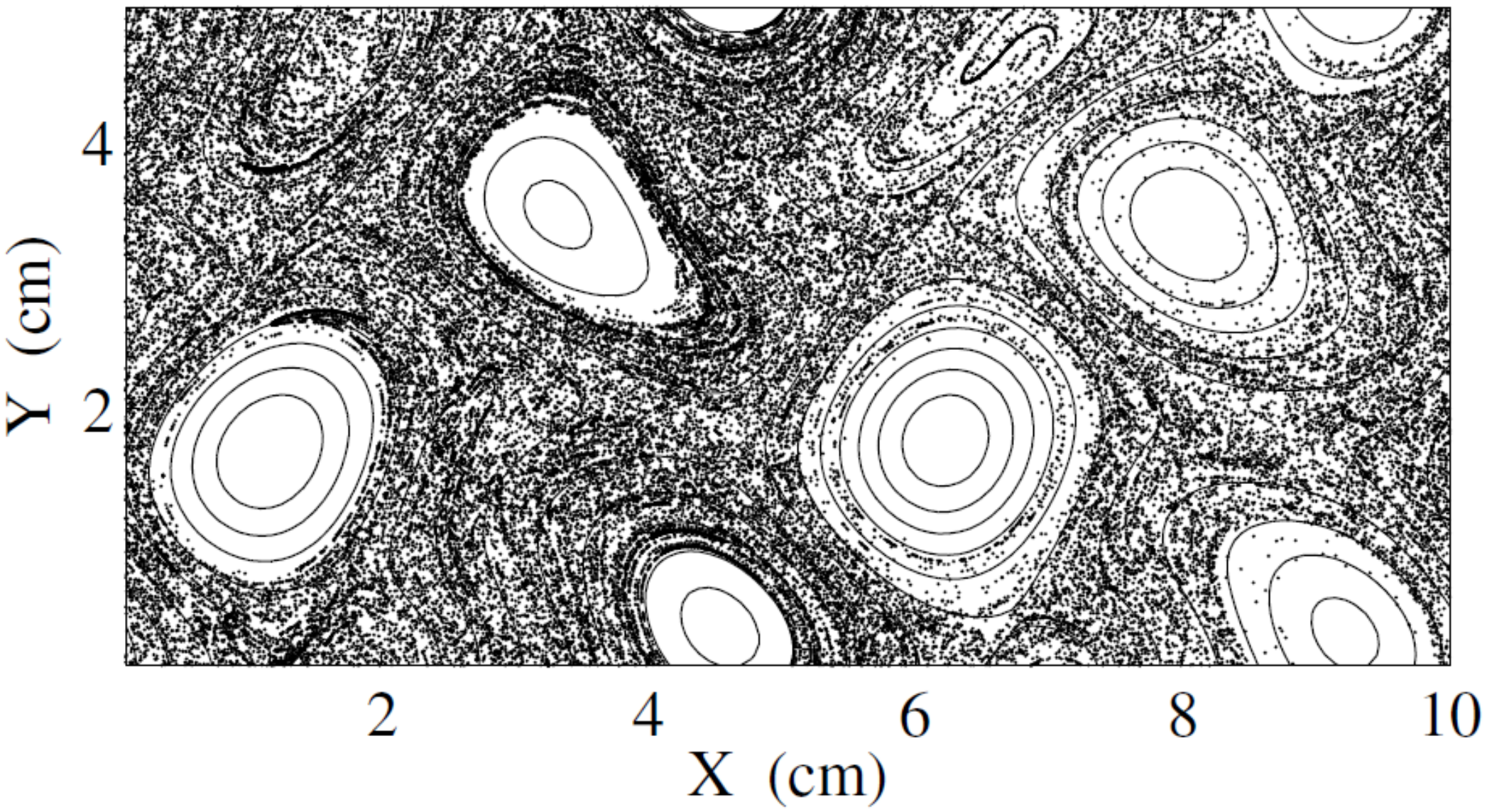}\hspace{5mm}
(b)
\includegraphics[width=3in]{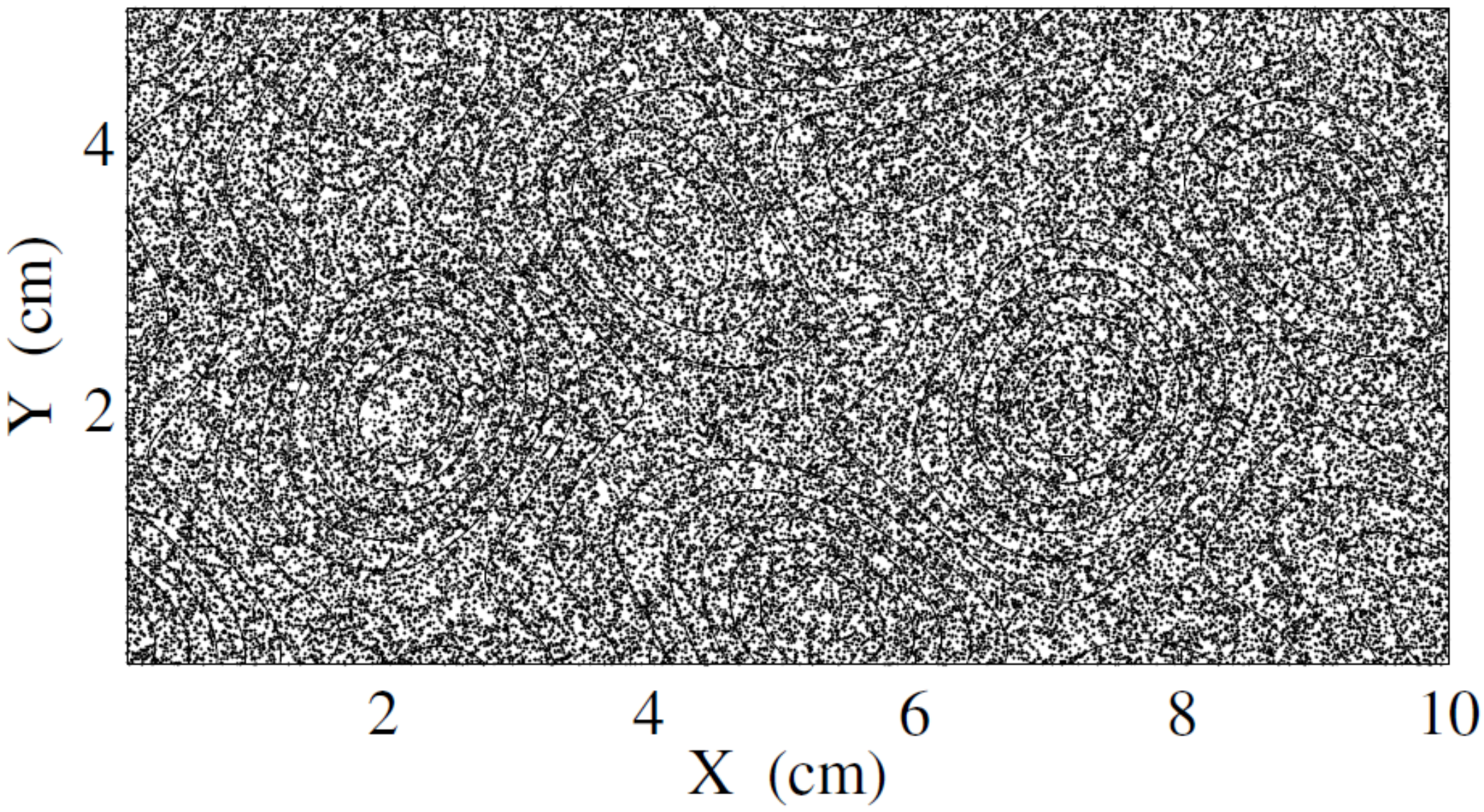}\\
(c)
\includegraphics[width=3in]{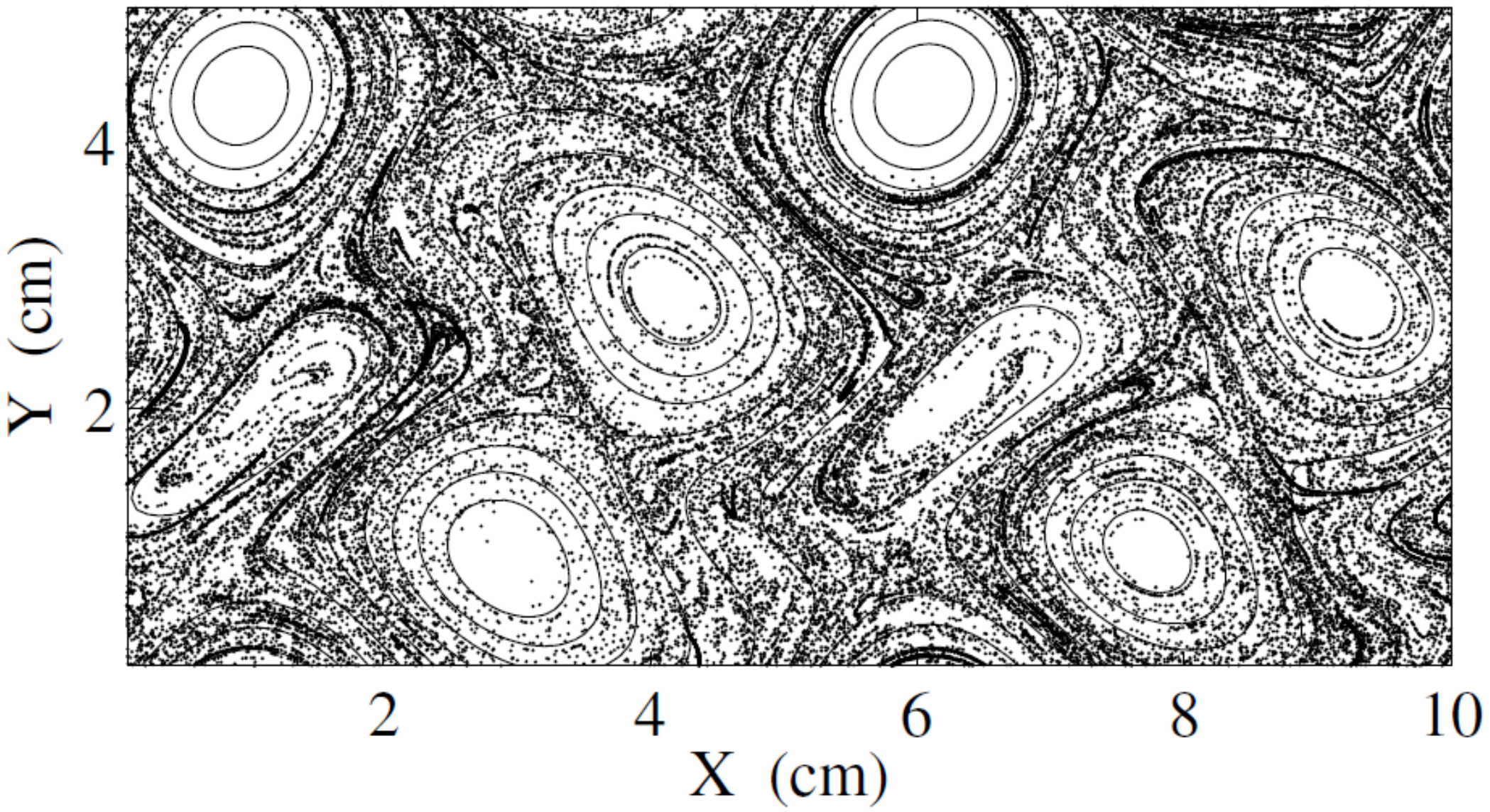}\hspace{5mm}
(d)
\includegraphics[width=3in]{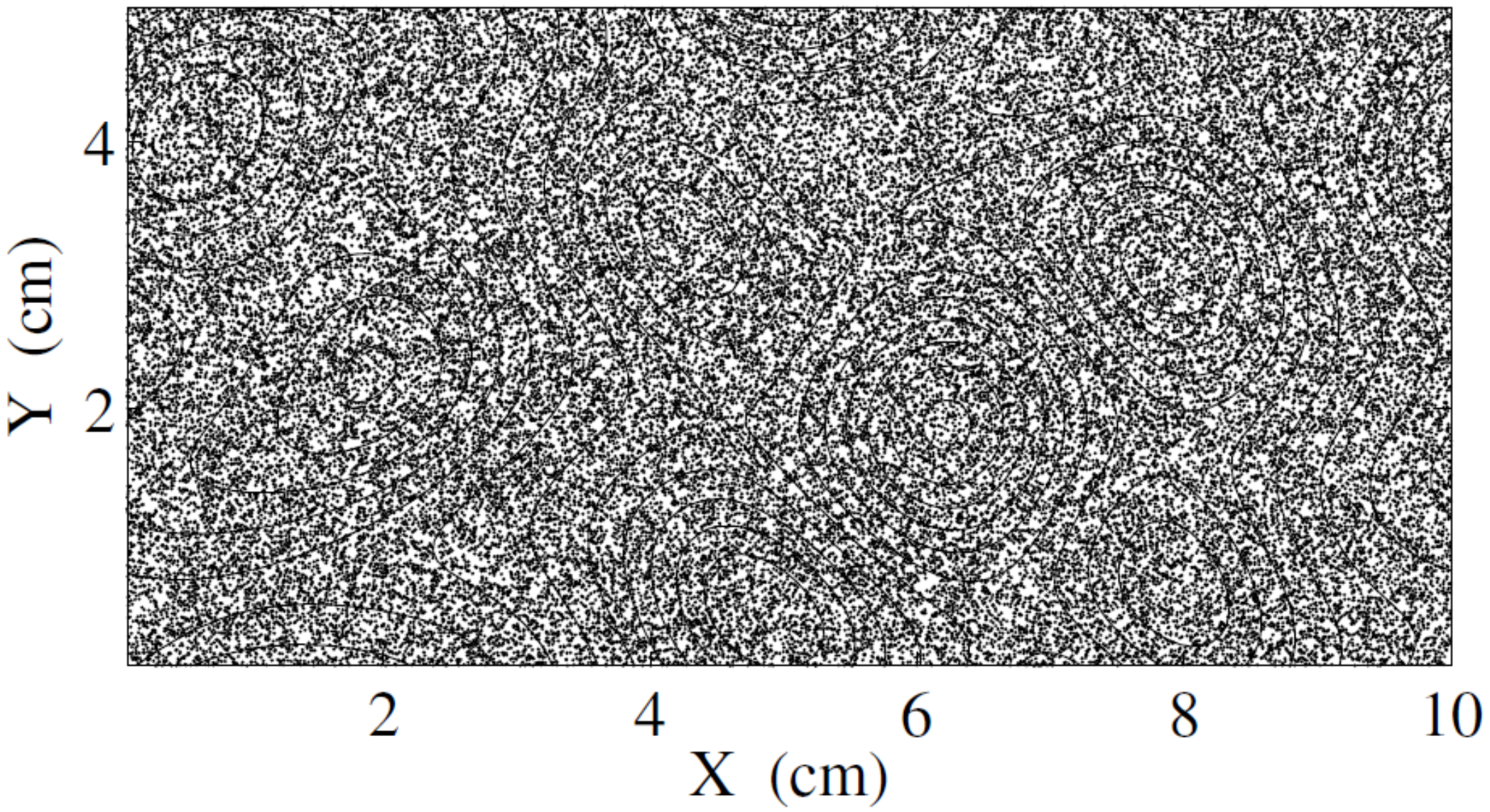}}
\caption{\label{tracerlevel} Mixing by aperiodic flow. Distribution of $6\times10^4$ tracers and the stream lines of the instantaneous flow for $A=0.872$ s$^{-2}$ at $t=369$ (a) and $t=3500$ s (b). The same for $A=0.878$ s$^{-2}$ at $t=221$ s (c) and $t=3500$ s (d).}
\end{figure*}

Fig. \ref{QP-mixing} shows the tracer distribution for two values of $A$ above the onset of the secondary Hopf bifurcation which destroys $P_2$ and makes the flow quasi-periodic. We find the evolution of the tracers to follow the same scenario as in the case of the time-periodic flow $P_3$: after a short initial stage of stretching and folding, the set of tracers fills a significant fraction of the full domain. This stage is followed by a much slower homogenization process in which the distribution becomes spatially uniform. However, just like in the case of $P_3$, the tracers never penetrate four regular islands centered around vortices, now with negative vorticity.

The fundamental difference between (quasi)periodic and aperiodic flows makes itself apparent if we compare mixing by the periodic flow $P_3$ with that by aperiodic flows just outside of the window of stability for $P_3$, at $A=0.872$ s$^{-2}$ and $A=0.878$ s$^{-2}$. Although the forcing is almost identical in these three cases and the short-term dynamics of the three flows are similar, Fig. \ref{tracerlevel} shows that the aperiodic flows achieve perfect mixing in the long term, covering the entire domain, including the four regular islands of $P_3$.

\begin{figure}[b]
\includegraphics[width=3in]{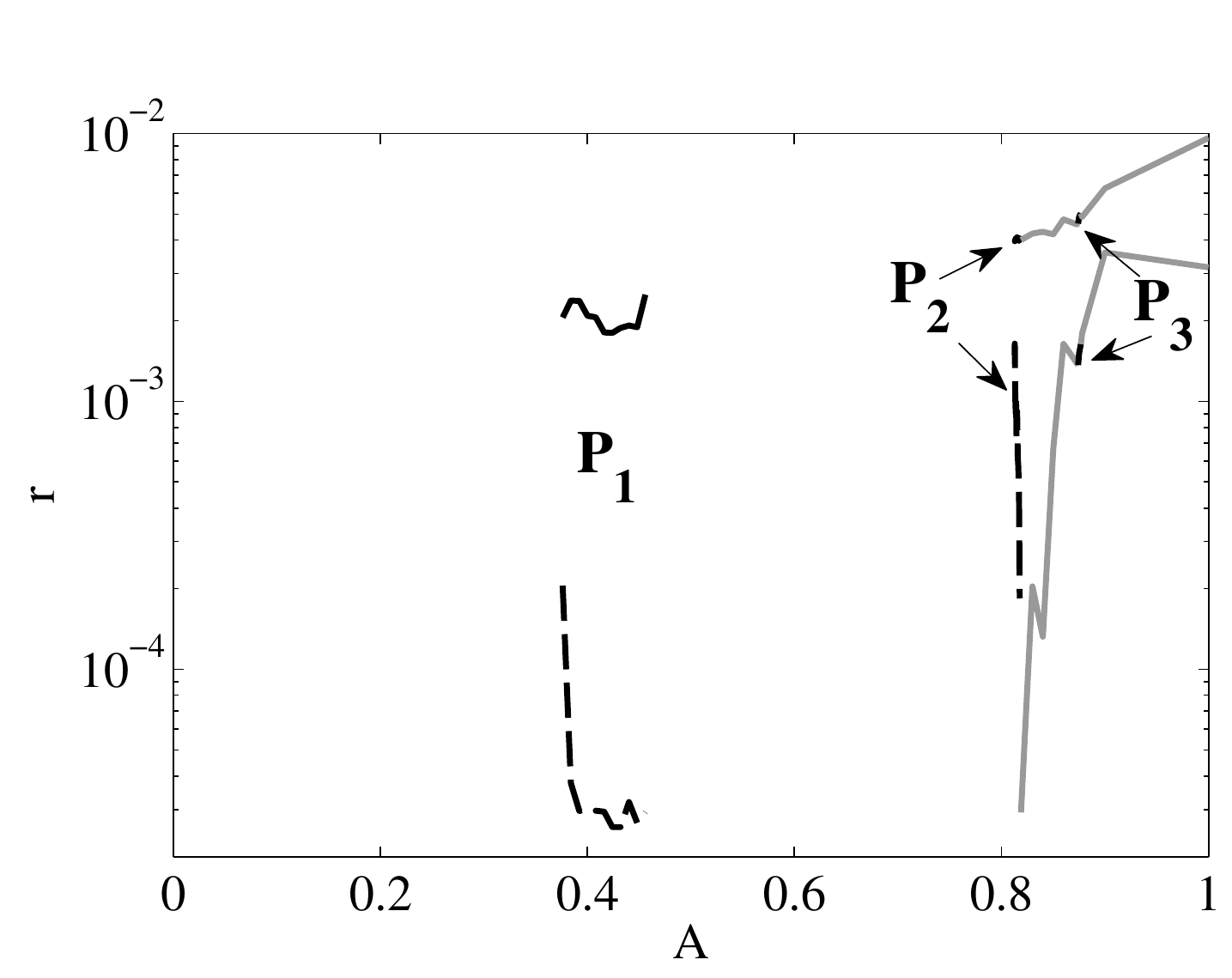}
\caption{The rates of mixing for time-dependent flows as a function of $A$. The solid and dashed curves correspond, respectively, the fastest time scale $r_{\max}$ and the slowest time scale $r_{\min}$.}
\label{fvst}
\end{figure}

Fig. \ref{fvst} summarizes the observed mixing rate as a function of the control parameter $A$. As Fig. \ref{fracvst} amply illustrates, the mixing process is characterized by a range of time scales. The fastest time scale describes stretching of the initial tracer distribution along the stream line passing through its center. The corresponding rate is defined as $r_{\max}=\max_t|df/dt|$  and is proportional to the average shear rate corresponding to that stream line.

The slowest time scale describes broadening of the distribution due to transport of tracers through semi-penetrable transport barriers discussed in Sect. \ref{s:resonance}. To characterize this broadening, we computed the time $t_{90}$ it takes for the area fraction to reach 90\% of its asymptotic value, $f(t_{90})/f(t_{100})=0.9$, where we assumed the asymptotic distribution is achieved at $t_{100}=5\times 10^4$ s. The minimal mixing rate was then defined as $r_{\min}=1/t_{90}$. In both cases we averaged $f(t)$ over a small window to filter out small oscillations associated with the passage of tracers near saddles.

The fast time scale $r_{\max}$ is found to increase almost monotonically with $A$, reflecting the corresponding increase in the shear of the underlying flow. The slow time scale requires more care to interpret. In particular, for $P_1$ we find $r_{\min}$ to drop by almost an order of magnitude as $A$ increases. This decline is associated with the tracer distribution shown in Fig. \ref{P12-mixing}(b) slowly broadening in time as illustrated by Fig. \ref{fracvst}(a). This broadening is due to a slow ``leak'' of tracers across a semi-penetrable transport barrier, creating a ``halo'' of tracers surrounding the main band. Another drop observed around the secondary Hopf bifurcation at $A\approx 0.818$ s$^{-2}$ is associated with a similar process for the quasi-periodic flow $QP$. As $A$ increases past this critical value of $A$, the transport barrier which exists for $P_2$ gets eroded, leading to a quick increase in $r_{\min}$. 

While many of our numerical results are quite logical, several findings raise questions. For instance, the flows $P_1$ and $P_3$ appear to be qualitatively quite similar. Both are stable, time-periodic and, with the choice of $A=0.428$ s$^{-2}$ for $P_1$, both have a time-dependent component of the same magnitude $\epsilon \approx 0.238$. Yet, despite these similarities, their mixing properties are radically different. $P_1$ is a very poor mixer, as Fig. \ref{P12-mixing}(b) illustrates. It is characterized by both a very low mixing rate and a very low mixed area fraction. In fact, $P_1$'s mixing properties are comparable to those of time-independent flows. $P_3$, on the other hand, is an extremely good mixer, almost as good as the aperiodic flows. The mixing rate for this flow is high and its mixed area fraction is close to unity.

Another question concerns the islands surrounding positive or negative vortices that remain impenetrable for extremely long times for both the time-periodic flow $P_3$ (Fig. \ref{P3-mixing}) and the quasi-periodic flow $QP$ succeeding $P_2$ (Fig. \ref{QP-mixing}). In both cases there appear to be transport barriers surrounding vortices characterized by vorticity of one sign but not the other. This was also found to occur in the model flow of Danilov {\em et al.} \cite{Danilov2000} as well as in real oceanic flows \cite{haller2011}.

\subsection{Lagrangian Coherent Structures}
\label{s:LCS}

\begin{figure}
\centering
(a)
\includegraphics[width=3in]{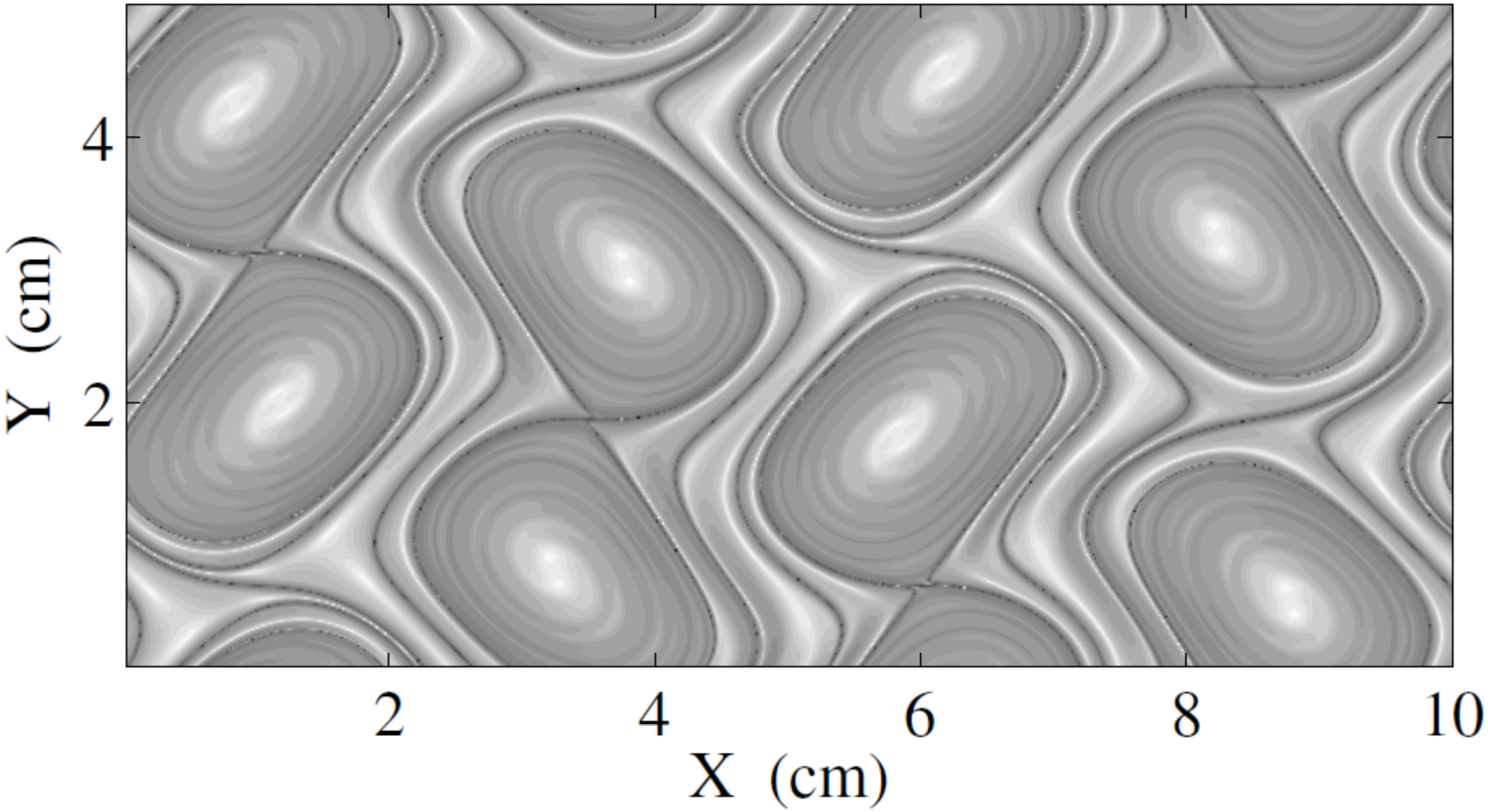}
\\
(b)
\includegraphics[width=3in]{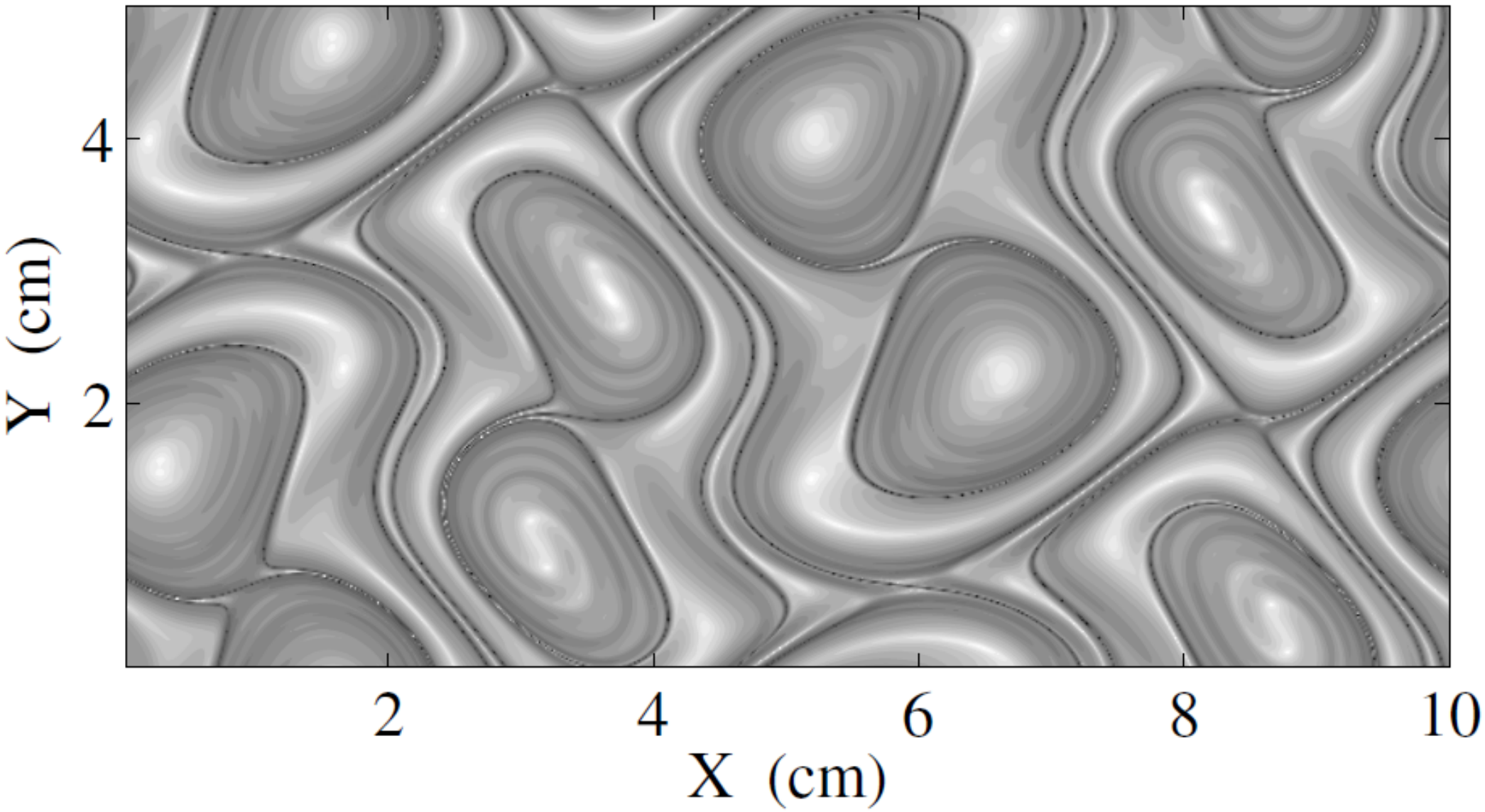}
\\
(c)
\includegraphics[width=3in]{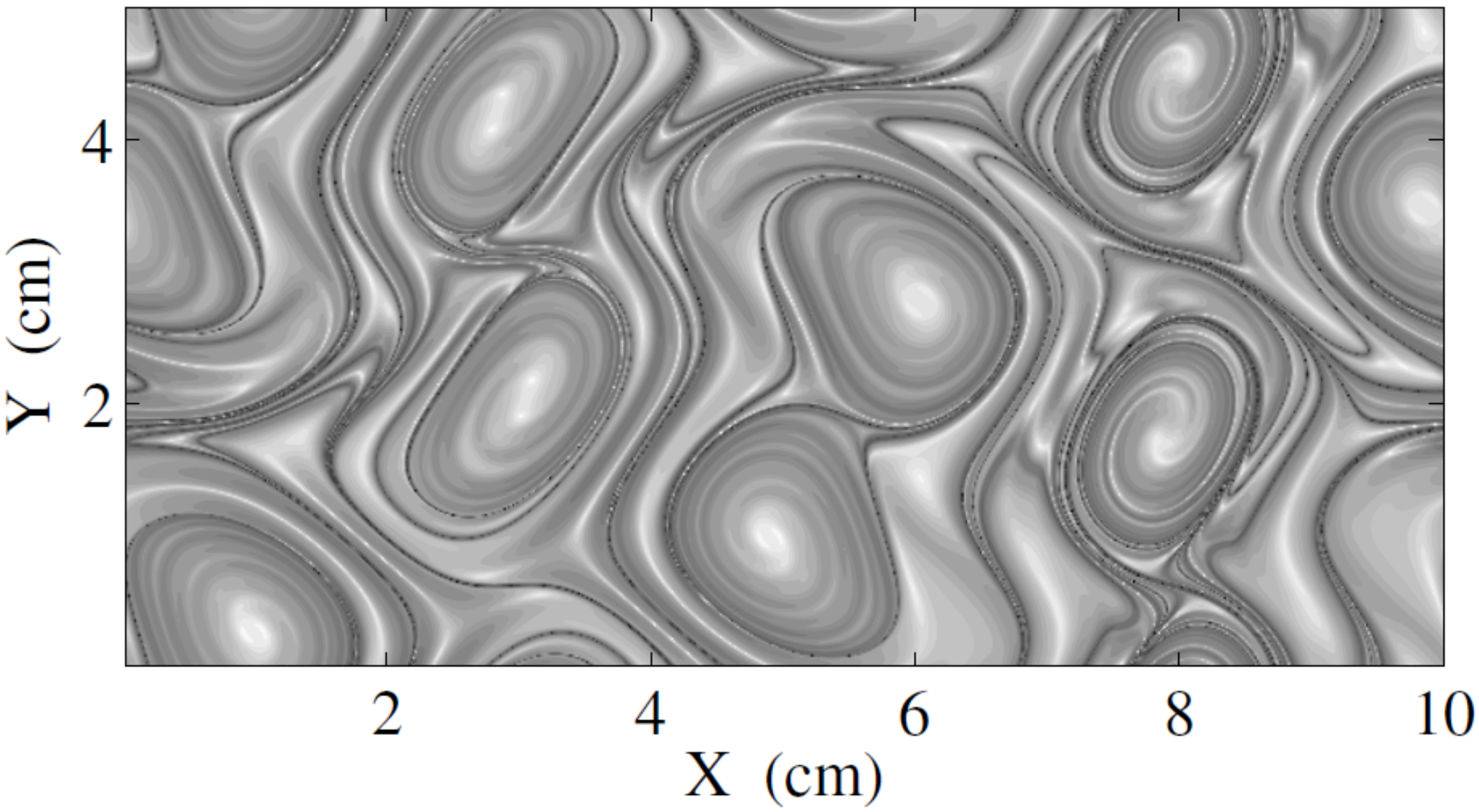}

\caption{\label{ftlefig} Forward finite-time Lyapunov exponent field. (a) $P_1$ at $A=0.428$ s$^{-2}$ with $\tau=32$ s, (b) $P_2$ at $A=0.817$ s$^{-2}$ with $\tau=19$ s, and (c) $P_3$ at $A=0.875$ s$^{-2}$ with $\tau=22$ s.}
\end{figure}

Finite-time Lyapunov exponent (FTLE) fields associated with the time-dependent flows provide some intuition regarding their drastically different mixing properties. The forward FTLE is a scalar quantity
\begin{equation}\label{ftle}
\sigma({\bf x}_0,\tau)=\frac{1}{\tau}\ln\left\|\frac{D{\bf x}(\tau)}{D{\bf x}_0}\right\|_2,
\end{equation}
which characterizes the amount of stretching along a trajectory ${\bf x}(t)$ passing through the point ${\bf x}_0$ at $t=0$ over a finite time interval $\tau$. In particular, the ridges of the forward FTLE field define Lagrangian Coherent Structures (LCS) \cite{shadden2005} which, for time-periodic flows, correspond to segments of unstable manifolds of saddle orbits with temporal period equal to that of the flow. As Fig. \ref{ftlefig} illustrates, for $P_1$ and $P_2$ the LCS show very little folding, effectively forming closed, compact curves. For $P_3$, on the other hand, the LCS display a lot of folding, which is a necessary ingredient for efficient mixing and cover a substantial fraction of the total area. Indeed, we find the LCS of $P_1$ and $P_2$ are qualitatively similar to those of steady flows from which they are born (i.e., $M$ and $N$), while the LCS of $P_3$ are qualitatively similar to those of aperiodic flows, which is consistent with the observed similarities in their mixing properties.

LCS play an important role in organizing transport. For instance, placing the initial set of tracers on top of the saddle orbit we should expect that set to be quickly stretched along the LCS forming effectively one-dimensional structures for $P_1$ and $P_2$, while for $P_3$ the structure becomes effectively two-dimensional. Furthermore, the LCS form transport barriers which cannot be crossed by the tracers. For $P_1$ and $P_2$ (as well for steady flows), these transport barriers are closed, effectively partitioning the domain and preventing mixing between regions separated by the LCS. For $P_3$ (as well as for aperiodic flows), the transport barriers are open, enabling transport and mixing across the whole domain.

The LCS-based description of transport is consistent with our long-term numerical advection calculations and has the advantage that it requires time-integration over a considerably shorter time-interval (fraction of the temporal period $T$ of the flow, compared with hundreds to thousands of periods for numerical advection calculations). However, neither approach explains {\em why} the mixing properties of the time-periodic flows are so dramatically different. A more insightful approach is discussed next.

\subsection{Resonance Phenomena}
\label{s:resonance}

As we mentioned previously, area-preserving time-periodic flows $P_1$, $P_2$, and $P_3$ can be treated formally as a perturbed Hamiltonian system (\ref{tracers}), with the stream function (\ref{perturbation}) serving the role of the Hamiltonian. In particular, $\Psi_0$ plays the role of the unperturbed Hamiltonian and $\epsilon\Psi_1$ -- the time-periodic perturbation. Transport in near-integrable time-periodic Hamiltonian systems and area-preserving flows has been studied extensively. It is well understood that, for weak perturbations, chaotic trajectories emerge in the neighborhood of the homo- or heteroclinic manifolds of saddle orbits of the integrable unperturbed, or base, flow. These manifolds self-intersect as a result of the imposed perturbation, forming a homoclinic tangle with the lobe dynamics \cite{Rom-Kedar1990} which provides an insightful, albeit computationally challenging, description of mixing in the separatrix chaotic layer (SCL).

Not only is the computation of the width of the SCL is a computationally intractable problem for any realistic flow, the width of the SCL significantly underestimates the size of the actual chaotic domain for finite values of $\epsilon$. According to the KAM theory \cite{kolmogorov1954,arnold1965,moser1962}, in the presence of perturbation, resonant tori of the unperturbed flow (tori whose frequency $\omega_0(\Psi_0)$ is in rational ratio with the frequency of the perturbation $\omega_1=2\pi/T$) break up, forming chains of elliptic and hyperbolic time-periodic orbits (or stream lines) with their own sets of self-intersecting stable and unstable manifolds generating resonant chaotic layers (RCL). These RCLs can overlap with the SCL making the chaotic domain much broader.

The dynamics away from the separatrix can be described by computing the change in the value of $\Psi_0$ over an interval of time $(0,t_f)$. By analogy with the derivation of Melnikov's function, we can use (\ref{perturbation}) and (\ref{tracers}) to show that
\begin{equation}\label{dpsi}
\Psi_0(t_f)-\Psi_0(0)=\epsilon\int_{0}^{t_f}v_0({\bf x}(t))v^\perp_1({\bf x}(t),t)dt,
\end{equation}
where ${\bf v}_i=(\partial_y\Psi_i,-\partial_x\Psi_i)$ and the superscript $\perp$ denotes the component normal to the stream line of the unperturbed flow. The velocity field describing a time-periodic perturbation can be written in the form of a Fourier series
\begin{equation}\label{fourier1}
v_1^\perp({\bf x},t)=\sum_kg({\bf x},\omega_k)e^{-i\omega_k t},
\end{equation}
where we have defined $\omega_k\equiv k\omega_1$. Furthermore, the product $v_0({\bf x}(t))g({\bf x},\omega_k)$ is also a time-periodic function with period $T_0=2\pi/\omega_0$ and can be written as a Fourier series
\begin{equation}\label{fourier2}
v_0({\bf x}(t))g({\bf x},\omega_k)=\sum_mG_{k,m}e^{im\omega_0 t}.
\end{equation}
Substituting (\ref{fourier1}) and (\ref{fourier2}) into (\ref{dpsi}) we find that, over a time interval $t_f\gg\max(T,T_0)$, the rate of change
\begin{equation}\label{rate}
\frac{\Psi_0(t_f)-\Psi_0(0)}{t_f}=\sum_{k,m} G_{k,m}\frac{\epsilon}{t_f}
\int_{0}^{t_f}e^{i(m\omega_0-\omega_k)t}dt
\end{equation}
vanishes unless $\omega_0/\omega_1=k/m$ for some integer $m$ and $k$. In what follows, we take $m$ and $k$ to be positive. 

\begin{figure}[t]
\hspace{2.5mm}\includegraphics[width=3.1in]{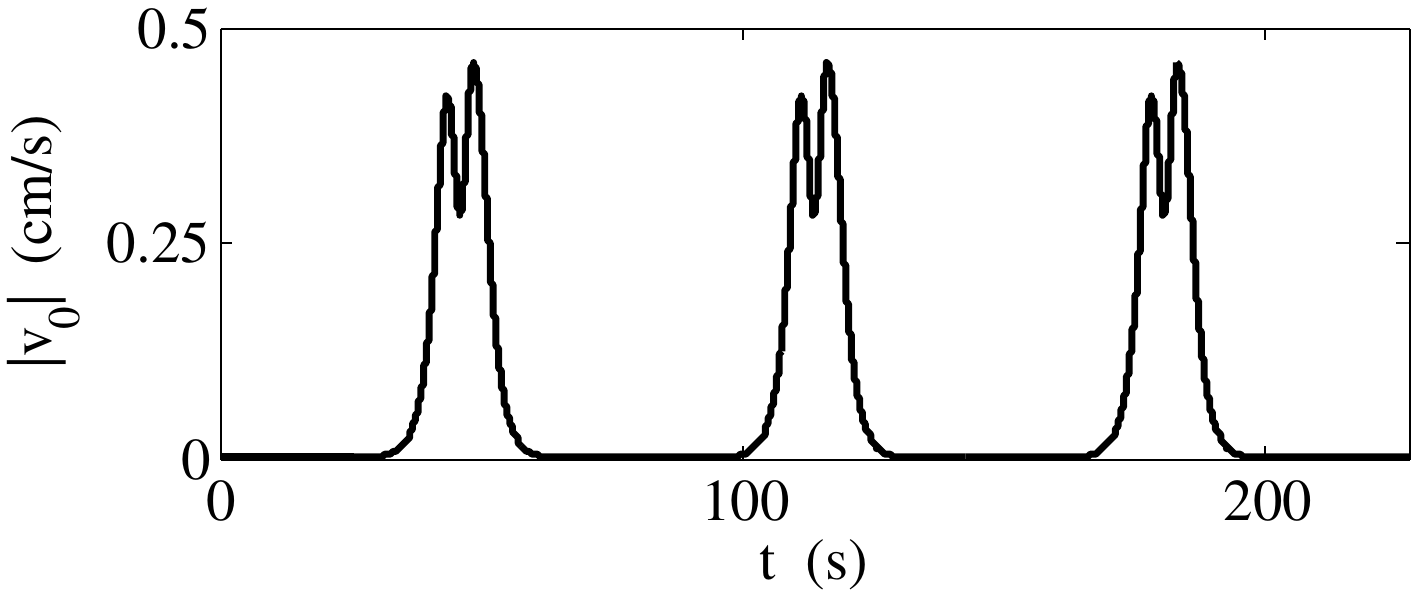}
\caption{Velocity magnitude for a typical stream line near a separatrix of the base flow of $P_1$.}
\label{velocity}
\end{figure}

The Fourier coefficient $G_{k,m}$ also controls the width $W^l_{k,m}$ of the RCL which replaces the stream line of the unperturbed flow with frequency $\omega_0=\omega_k/m$. For stream lines close to the separatrix $\Psi_0(x,y)=\psi_l$, $v_0({\bf x}(t))$ is small everywhere except for short time intervals corresponding to fast motion away from the saddles (see Fig. \ref{velocity}), so we can estimate
\begin{eqnarray}\label{fourier3}
|G_{k,m}|&=&\left|\frac{1}{T_0}\int_0^{T_0} v_0({\bf x}(t))g({\bf x}(t),\omega_k)
e^{-im\omega_0 t}dt\right|\nonumber\\
&\approx&\frac{s_l}{2\pi}\frac{\omega_k|g({\bf x}_l^*,\omega_k)|}{m},
\end{eqnarray}
where $s_l$ is the length of the separatrix,
\begin{equation}
g({\bf x}_l^*,\omega_k)=\frac{1}{T}\int_0^{T} v_1^\perp({\bf x}^*,t)e^{i\omega_k t}
\end{equation}
is the (discrete) Fourier spectrum of $v_1^\perp({\bf x}^*,t)$, and ${\bf x}_l^*$ is the point on the separatrix which lies midway between the saddles (for which $v_0({\bf x}(t))$ is near its maximum). If there is more than one saddle on the separatrix, the estimate (\ref{fourier3}) should be generalized to include respective contributions from all segments, which can either enhance or suppress each other.

\begin{figure*}
\centering
(a)\hspace{-5mm}
\includegraphics[width=3in]{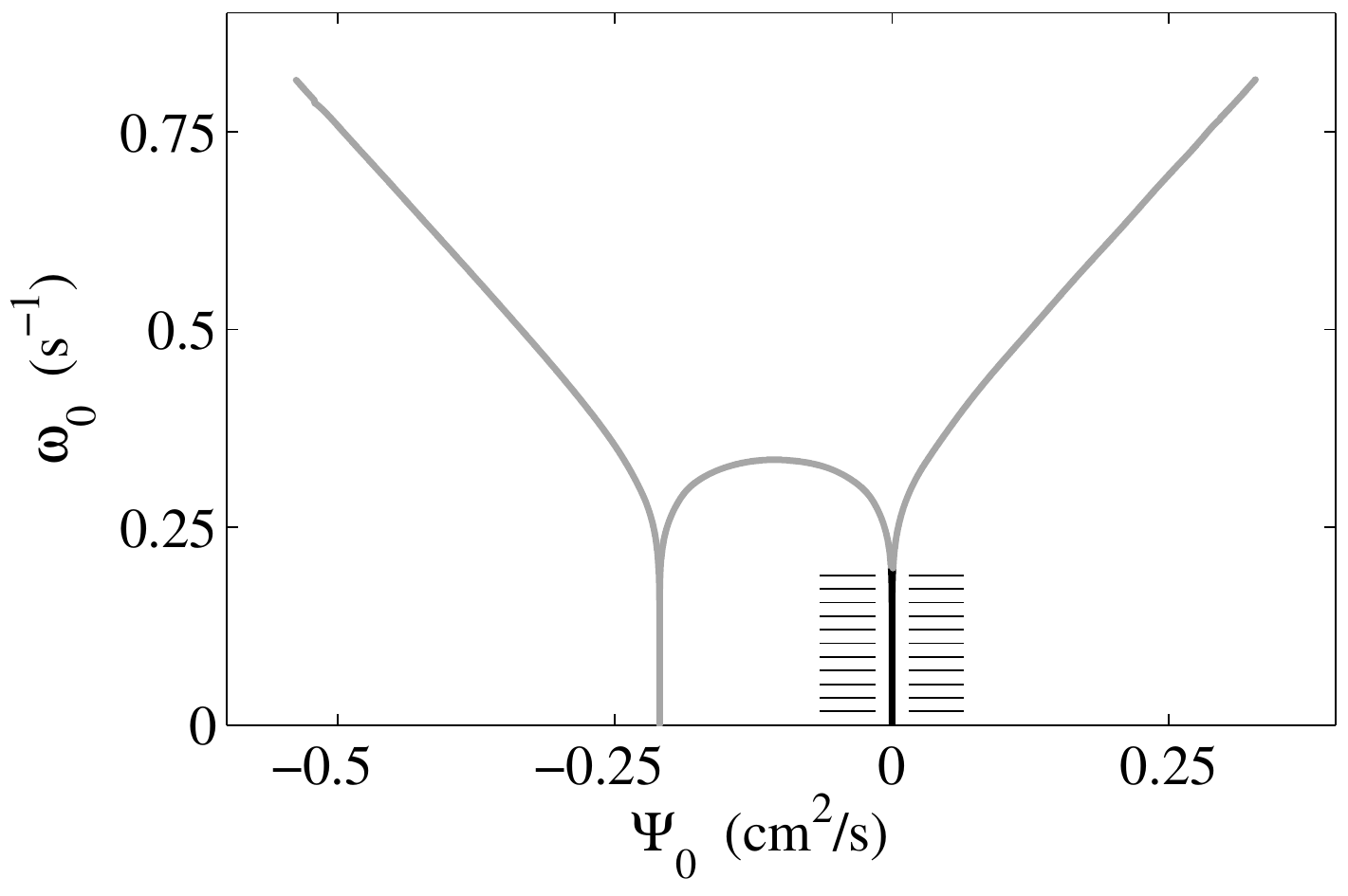}\hspace{5mm}
(b)\hspace{-5mm}
\includegraphics[width=3in]{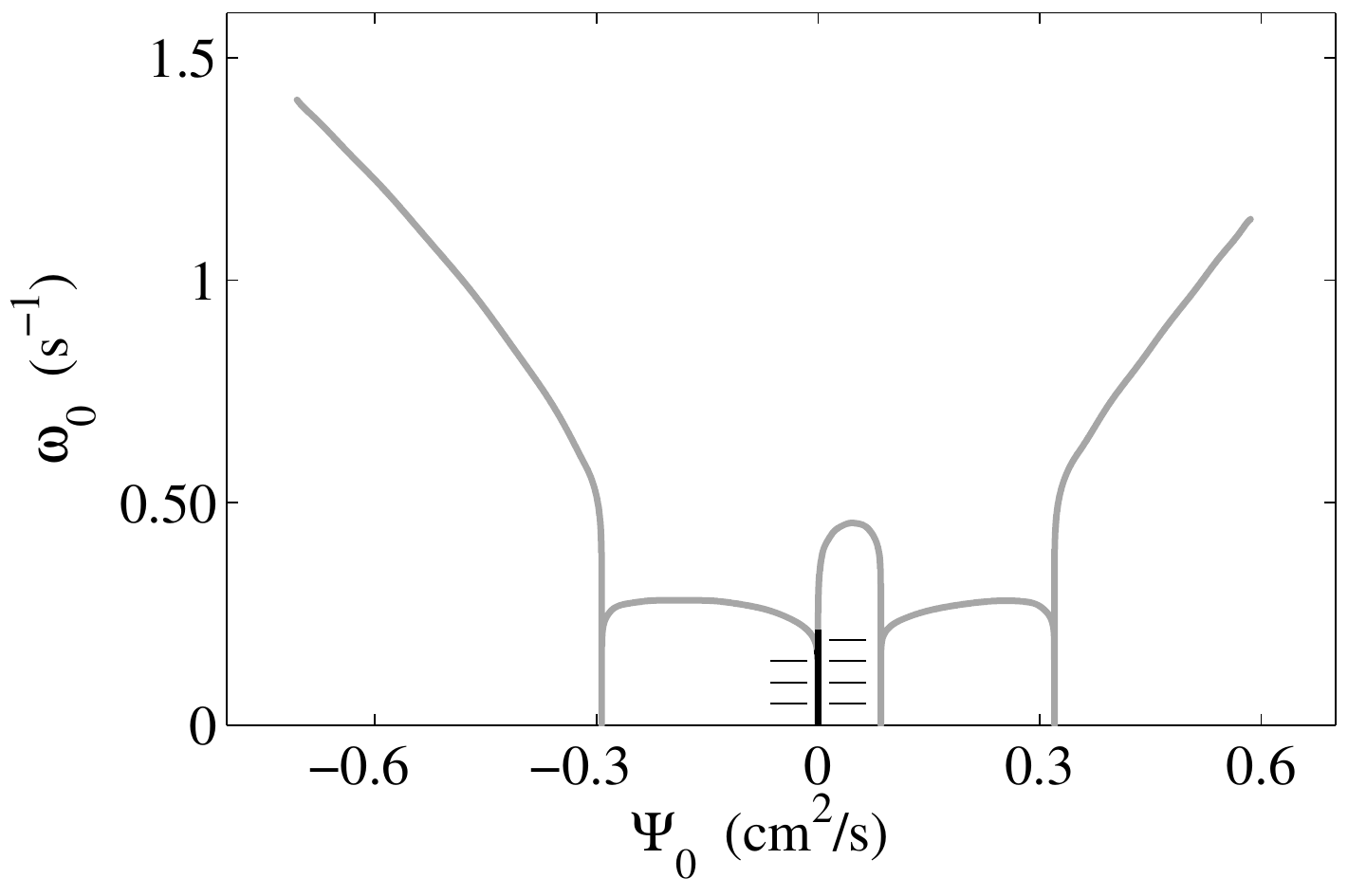}
\\
(c)\hspace{-5mm}
\includegraphics[width=3in]{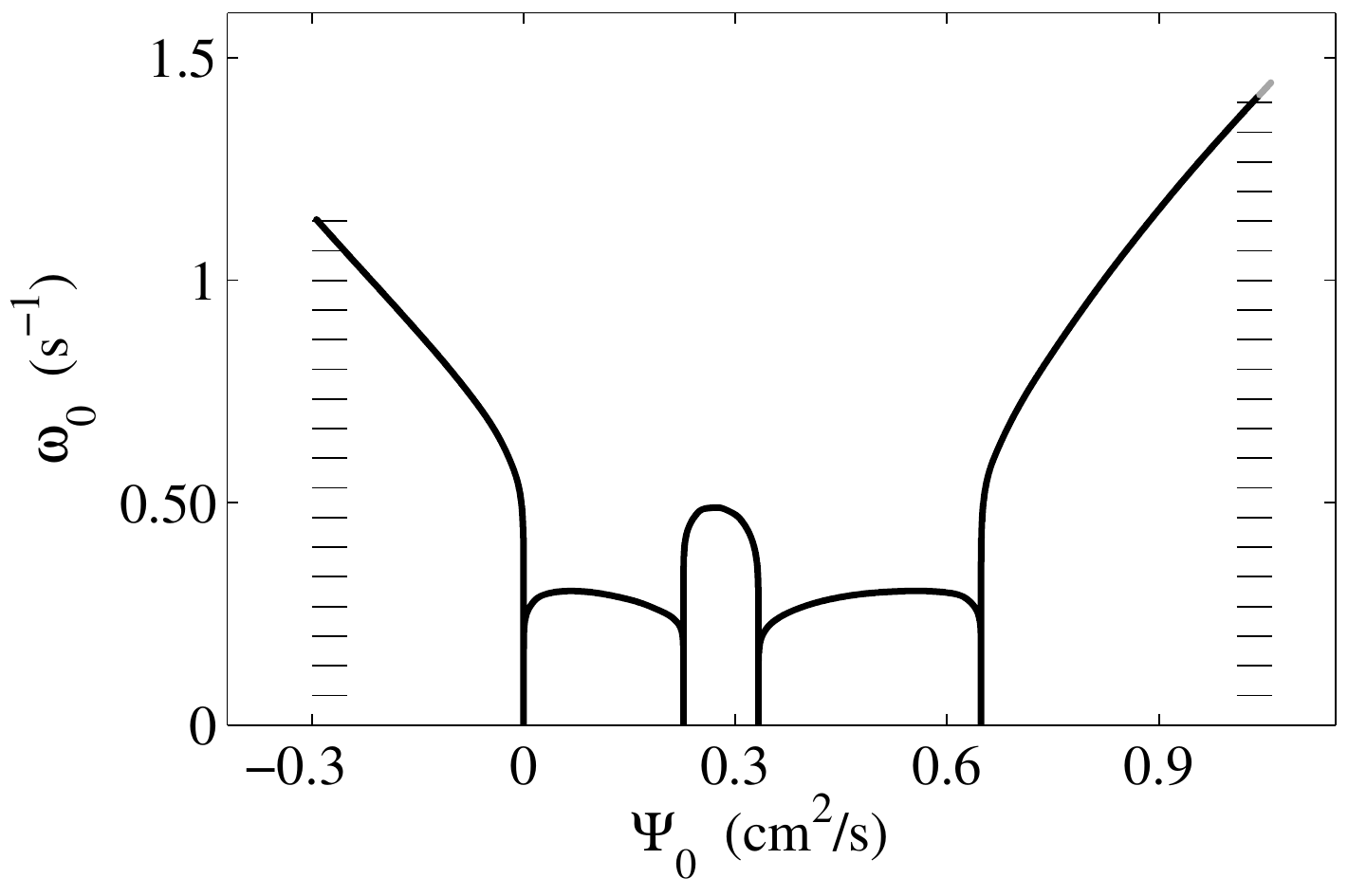}\hspace{5mm}
(d)\hspace{-5mm}
\includegraphics[width=3in]{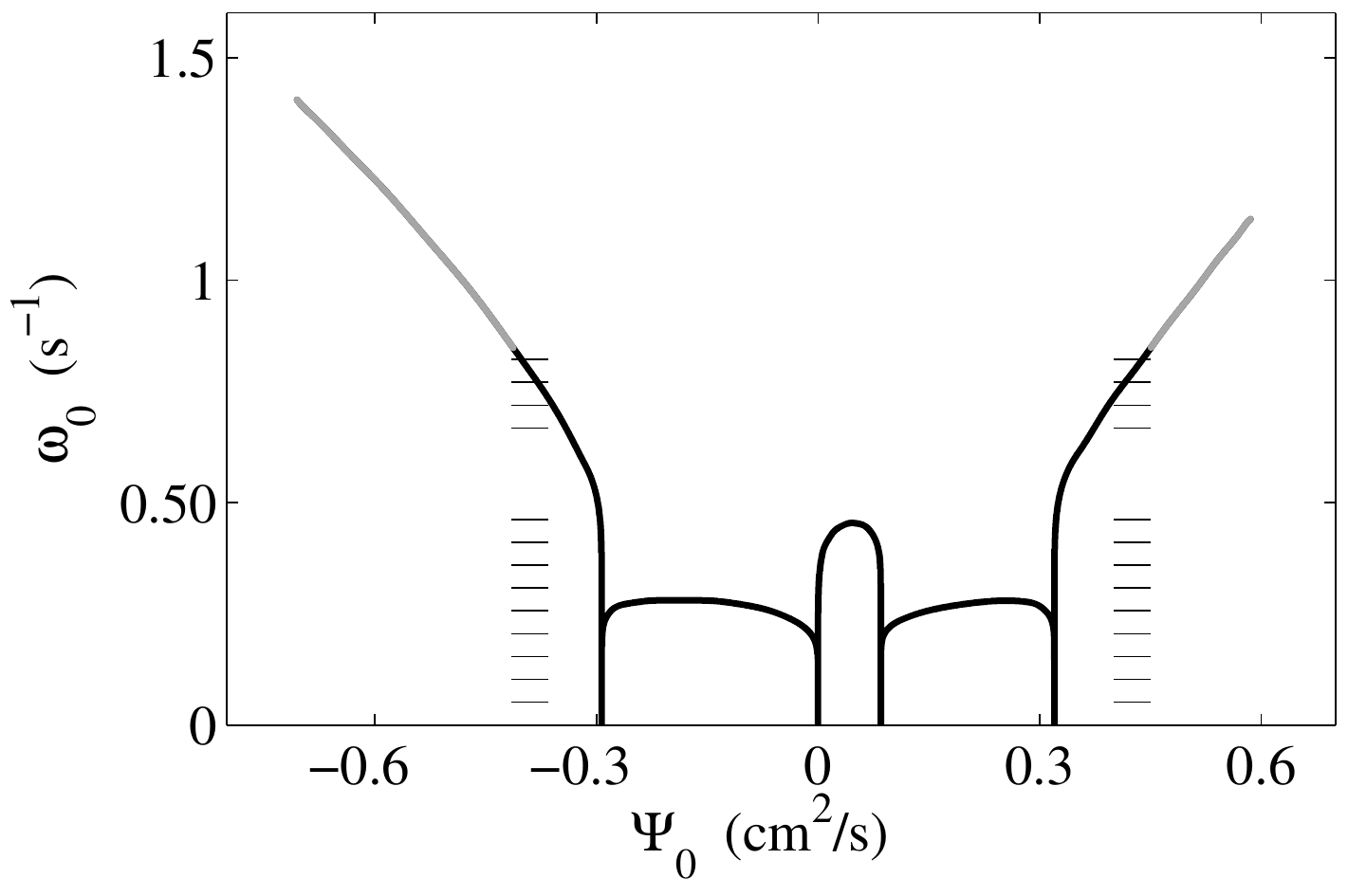}
\caption{\label{ompsi} Frequency of the motion along the stream lines of the base flow for (a) $P_1$ at $A=0.428$ s$^{-2}$, (b) $P_2$ at $A=0.817$ s$^{-2}$, (c) $P_3$ at $A=0.875$ s$^{-2}$ and (d) $QP$ at $A=0.820$ s$^{-2}$. The black portion corresponds to the chaotic domain around the separatrix $\Psi_0=0$ seeded with the tracers, while the gray portion corresponds to the regular (as well as chaotic) domains without tracers. Horizontal lines show the frequencies of the overlapping dominant RCLs.}
\end{figure*}

The period of the unperturbed motion along the stream line lying near the same separatrix is given by
\begin{equation}\label{period}
T_0(\Psi_0)=-\sum_i\lambda_{l,i}^{-1}\ln\frac{|\Psi_0-\psi_l|}{\xi_l},
\end{equation}
where $\lambda_{l,i}$ are the positive eigenvalues of all the saddles on the separatrix and $\xi_l$ is a constant. Hence, the distance (in terms of $\Psi_0$) from the separatrix to the nearest $k$:$m$ resonant torus is exponentially small for low $k$:
\begin{equation}
|\Psi_0-\psi_l|=\xi_l\exp\left(-\frac{\chi_l}{\omega_0}\right)=\xi_l\exp\left(-\frac{m\chi_l}{k\omega_1}\right),
\end{equation}
where $\chi_l^{-1}\equiv\sum_i\lambda_{l,i}^{-1}/2\pi$.
In a similar fashion we can compute the distance between various resonant tori. In particular, the distance between the tori with frequency ratios $k$:1 and $(k+1)$:1 is given by
\begin{equation}
S_l(\omega_k)\approx\left|\frac{d\Psi_0}{d\omega_0}\right|\omega_1\approx \frac{\xi_l\chi_l\omega_1}{\omega_k^2}\exp\left(-\frac{\chi_l}{\omega_k}\right).
\end{equation}

According to (\ref{fourier3}), for moderate $k$, $|G_{k,m}|$ takes the largest values for $m=1$, hence the width of the dominant ($k$:1) RCLs can be estimated from (\ref{rate}):
\begin{equation}
W_l(\omega_k)\approx\epsilon \frac{T}{2} |G_{k,1}|\approx\frac{\epsilon s_l}{2}\frac{\omega_k}{\omega_1}|g({\bf x}_l^*,\omega_k)|.
\end{equation}
Comparing the widths of the RCLs with the distances $S_l(\omega_k)$ between the neighboring resonant tori we can determine which RCLs overlap and which do not for a particular strength of the perturbation. 

For moderate $\epsilon$, we can expect several RCLs with low $k$ to overlap with each other and with the SCL, since $S_l(\omega_k)$ is exponentially small, while $W_l(\omega_k)$ scales as a power of $\omega_k$ near a separatrix. More specifically, the region where $W_l(\omega_k)>S_l(\omega_k)$ is expected to be well mixed, while in the region where $W_l(\omega_k)<S_l(\omega_k)$ mixing is expected be limited to narrow RCLs of width $W_l(\omega_k)$ (as well as some even narrower RCLs corresponding to $m>1$). The boundaries of the main chaotic region should be determined by the regular tori with frequency
\begin{equation}
\omega_0(\Psi_0(x,y))\approx \omega_{k_\pm}+\frac{W_l(\omega_{k_\pm})}{2}\left|\frac{d\omega_0}{d\Psi_0}\right|_{\omega_{k_\pm}}.
\end{equation}
The value $k_\pm$ on each side is different and corresponds to the outermost overlapping RCL, i.e., it is largest integer $k_\pm$ such that $S_l(\omega_k)<W_l(\omega_k)$ for all $k<k_\pm$.

\begin{figure*}
\centering
(a)\hspace{-5mm}
\includegraphics[width=3in]{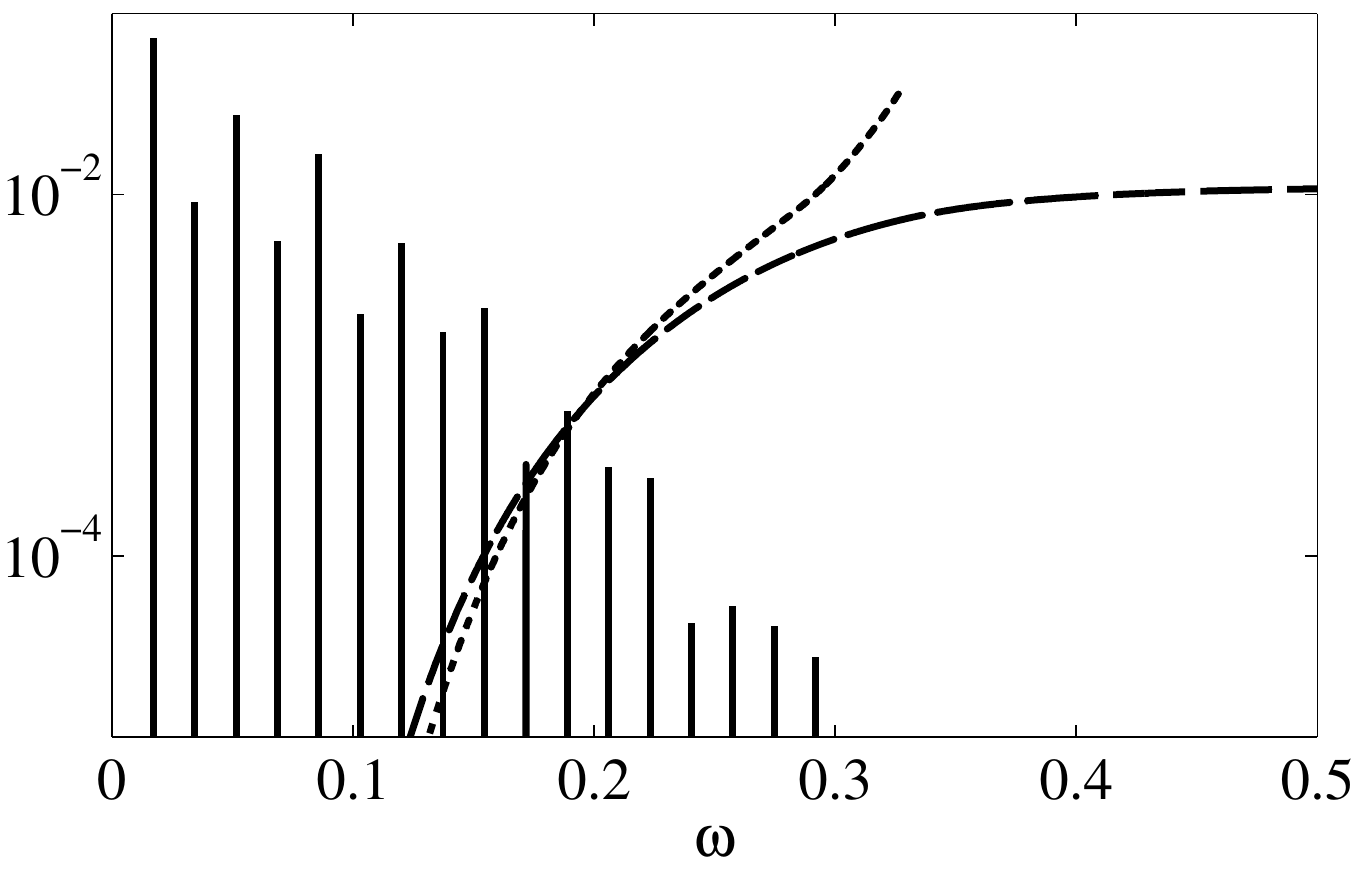}\hspace{5mm}
(b)\hspace{-5mm}
\includegraphics[width=3in]{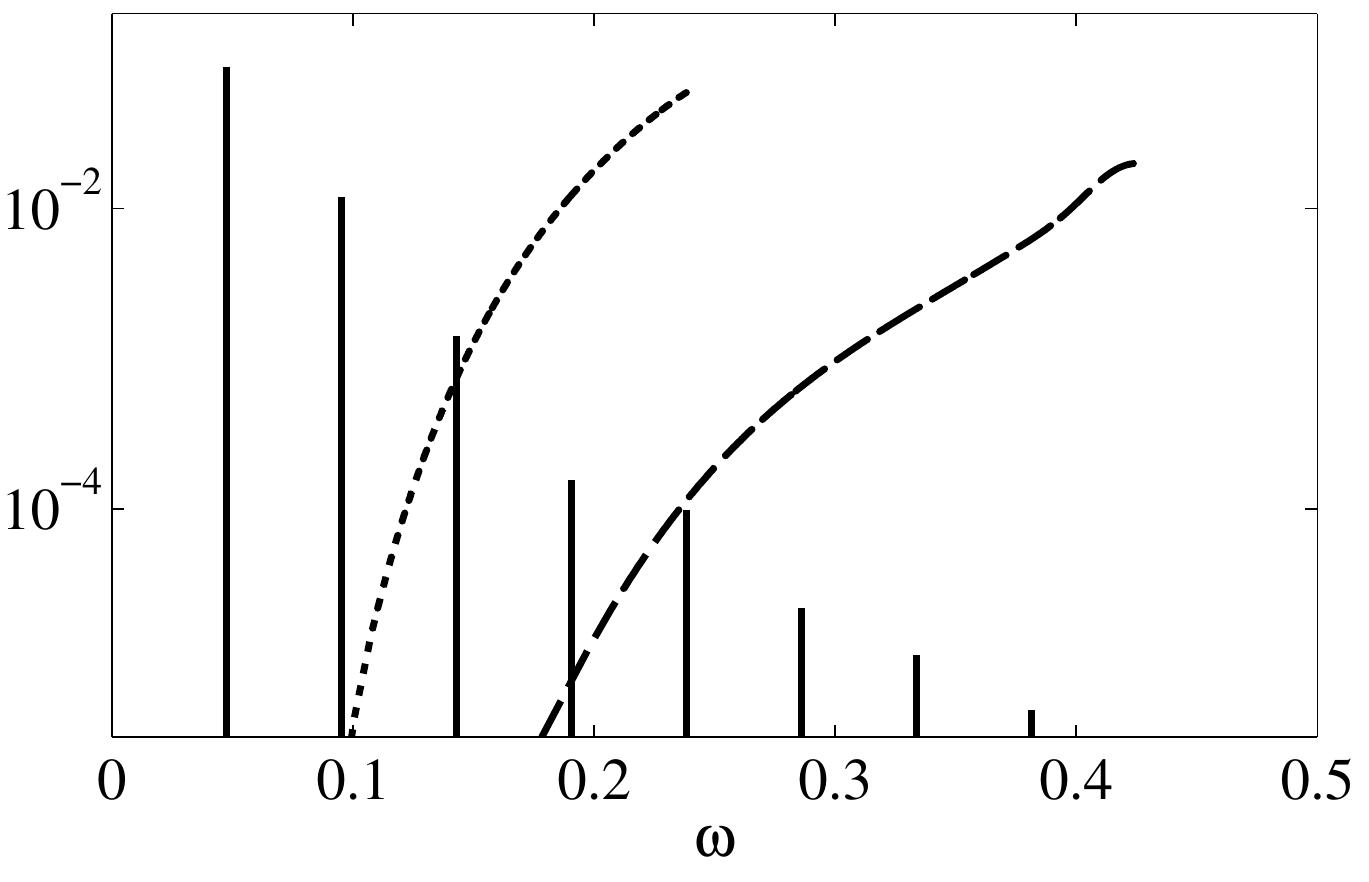}
\\
(c)\hspace{-5mm}
\includegraphics[width=3in]{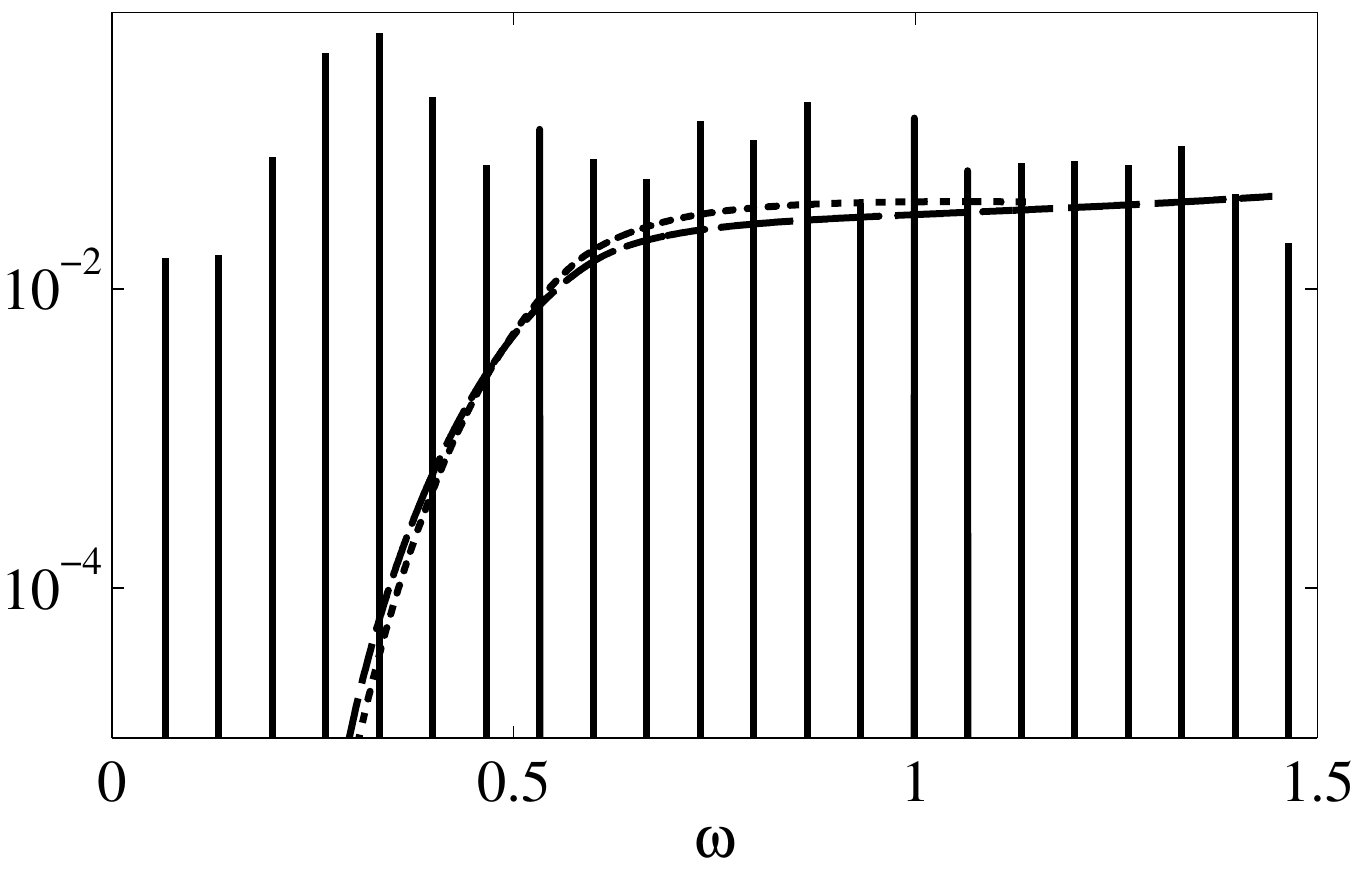}\hspace{5mm}
(d)\hspace{-5mm}
\includegraphics[width=3in]{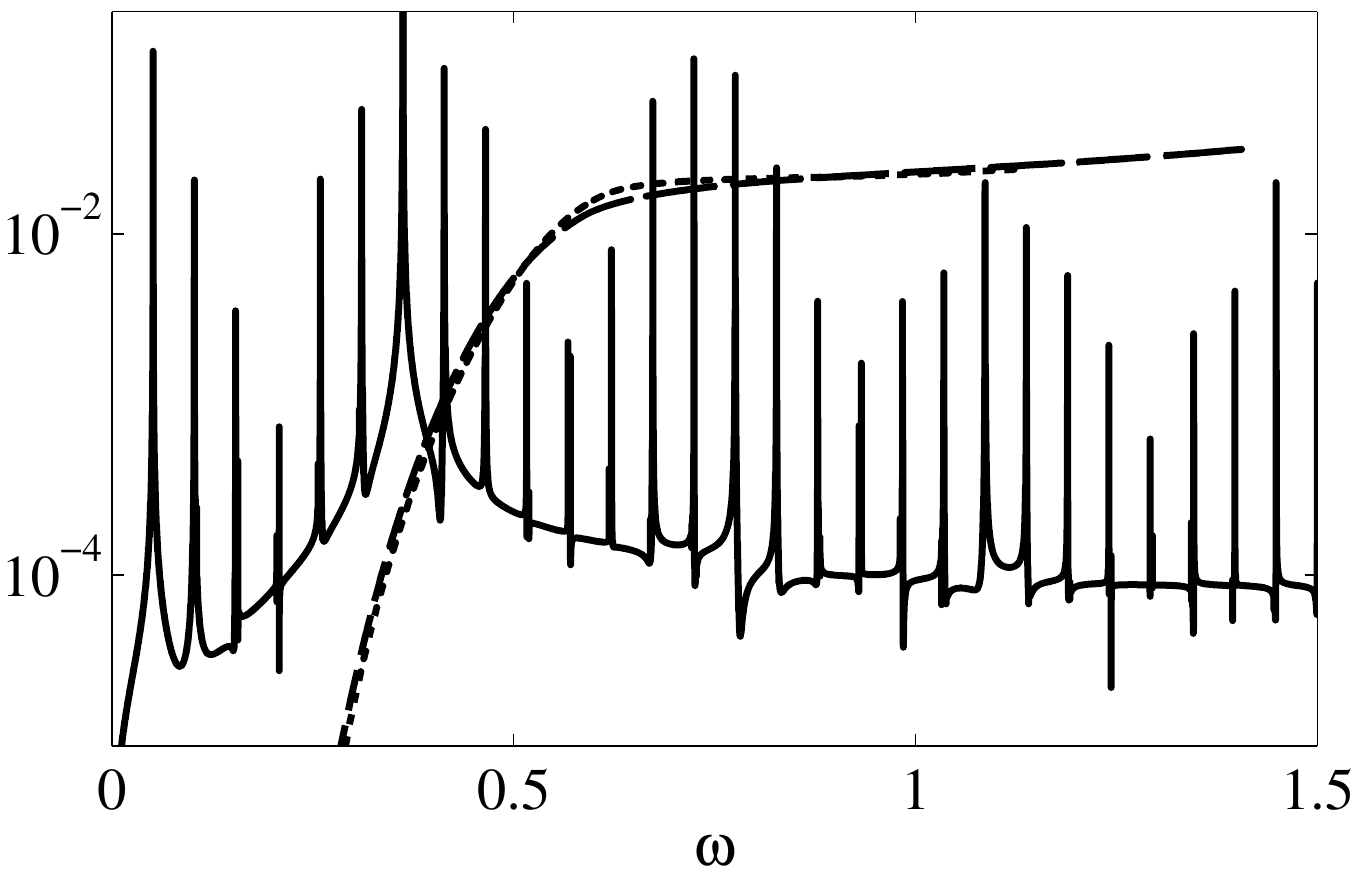}
\caption{\label{pwrfreq} The widths $W_l(\omega)$ of resonant chaotic layers (solid line) and the spacings $S_l(\omega)$ between the corresponding resonant tori (dashed and dotted lines) on different sides of the separatrix $\Psi_0=0$. (a) $P_1$ at $A=0.428$ s$^{-2}$ with $\omega_1=0.0172$ s$^{-1}$, (b) $P_2$ at $A=0.817$ s$^{-2}$ with $\omega_1=0.0477$ s$^{-1}$, (c) $P_3$ at $A=0.875$ s$^{-2}$ with $\omega_1=0.0666$ s$^{-1}$, (d) $QP$ at $A=0.820$ s$^{-2}$ with $\omega_1\approx0.051$ s$^{-1}$}
\end{figure*}

Consider, for example, the flow $P_2$. Fig. \ref{pwrfreq}(b) shows the width of resonant chaotic layers and the spacing between the resonant tori on both sides of the saddle seeded with tracers (which corresponds to $\Psi_0=0$). For low values of $k$ we do indeed find $W_l(\omega_k)>S_l(\omega_k)$, so the separatrix chaotic layer and the RCLs of a few nearby $k$:1 resonant tori overlap, forming a single chaotic domain. $W_l(\omega_k)$ decreases quickly (in fact, exponentially fast) with $\omega_k$, while $S_l(\omega_k)$ increases on both sides of the separatrix, so that the number of overlapping RCLs is rather low ($k_-=3$ for $\Psi_0<0$ and $k_+=4$ for $\Psi>0$), making the width of the chaotic domain (the black region in Fig. \ref{ompsi}(b)) extremely small, in good agreement with the numerical result shown in Fig. \ref{P12-mixing}(d).

The situation is similar for $P_1$. Fig. \ref{pwrfreq}(a) shows that the number of overlapping RCLs is somewhat larger for $P_1$ ($k_-=k_+=11$), but still small enough for the chaotic domain formed by the overlapping resonant and separatrix chaotic layers to remain quite thin (see Fig. \ref{ompsi}(a)). This is again in agreement with the numerical result shown in Fig. \ref{P12-mixing}(b). A ``halo'' of tracers outside of the well mixed region suggests that the outermost RCL just touches its neighbor on the side of the separatrix, with the two separated by a semi-penetrable transport barrier formed by either a narrow high-order RCL with very small $G_{k,m}$ or by a cantorus \cite{mackay1984}, which allows a very slow ``leak'' of tracers into the outermost RCL.

While $P_1$ and $P_2$ are almost monochromatic, the Fourier spectrum $g({\bf x}^*,\omega_k)$ of $P_3$ is exceptionally broad, with a large number of harmonics $\omega_k$ that have amplitudes comparable to that of the base frequency $\omega_1$. As a result, we find that RCLs remain fairly broad even for large values of $k$ (see Fig. \ref{pwrfreq}(c)). Consequently, the chaotic domain for $P_3$ is comprised of a large number of overlapping RCLs which allow transport across most of the $\Psi_0$ range (that is across most of the physical space). Since in this case the boundaries of the chaotic domain are very far from the separatrix of the saddle seeded with tracers, we computed the spacing $S_l(\omega_k)$ between the resonant tori corresponding to the two outermost branches of the $\omega_0(\Psi_0)$ curve which extend to the extremal values of $\Psi_0$ corresponding to clockwise and counterclockwise vortices. As Fig. \ref{ompsi}(c) illustrates, $W_l(\omega_k)>S_l(\omega_k)$ for all $k$  for both the leftmost and the rightmost branch. Hence, we should expect all of the RCLs to overlap, allowing global transport. However, the last RCL surrounding the clockwise vortex is not wide enough to cover the whole range of $\Psi_0$, leaving a small regular island around each of the four vortices centered at $\Psi_0(x,y)=1.058$, which is also in agreement with the numerical result shown in Fig. \ref{P3-mixing}(b).

The mixing properties of the quasi-periodic flow $QP$ can also be understood by analyzing the widths of RCLs. The time-average flow for $QP$ is essentially the same as that for $P_2$, hence we can use the same frequency curve $\omega_0(\Psi_0)$. The spectrum $g({\bf x}^*,\omega)$ of the quasi-periodic perturbation is discrete, just like the spectra of the periodic flows, with frequencies of the peaks that can be labeled $\omega_k$, with integer $k$. Although the spacing $\omega_{k+1}-\omega_k$ between the peaks is not exactly constant, it varies little about the average $\omega_1\approx 0.051$ s$^{-1}$, as Fig. \ref{pwrfreq}(d) illustrates. Comparison of $S_l(\omega_k)$ and $W_l(\omega_k)$ shows that RCLs with $k\le 9$ overlap. In addition, the tori 13:1, 14:1, 15:1, and 16:1 also overlap. Although the width of the tori 10:1, 11:1, and 12:1 is smaller than the spacing between them, the tori 9:1 and 13:1 are more than twice as wide as the biggest inter-tori spacing in this ``gap'', which means that the 9:1 and 13:1 RCLs overlap directly. This indicates that we should have global transport in the region of $\Psi_0$ where $\omega_0(\Psi_0)<\omega_{16}+W_l(\omega_{16})|d\omega_0/d\Psi_0|/2\approx 0.85$. According to Fig. \ref{ompsi}(d), this corresponds to almost the entire physical domain, with the exception of regular islands around all the vortices.

Although this prediction is {\em not} in perfect agreement with the numerical result, the description in terms of interacting resonances captures most of the features of the asymptotic tracer distribution shown in Fig. \ref{QP-mixing}(b). Both global transport and the regular islands around the clockwise vortices are predicted correctly. The size of the regular islands is predicted to be much larger than for $P_3$, which is also consistent with numerics. The resonant description also predicts the formation of regular islands around the counterclockwise vortices, which are filled in in Fig. \ref{QP-mixing}, which shows a limitation of our analytical description.

However, the discrepancies at high frequencies are expected, given the fact that the expression (\ref{fourier3}) for the Fourier coefficients $G_{k,m}$ (and hence the widths $W_l(\omega_k)$) was obtained in the limit of low frequencies $\omega_k$. For higher frequencies corresponding to the neighborhood of the vortices, (\ref{fourier3}) becomes inaccurate and $G_{k,m}$ has to be computed in a different way.

\section{Summary and Conclusion}
\label{s:summary}

To summarize, we have described transition from the laminar Kolmogorov flow to turbulence in a doubly-periodic domain of relatively small size. The sequence of bifurcations preceding turbulence is quite rich, with several different steady and time-periodic flows succeeding one another. This bifurcation sequence is quite sensitive to the choice of parameters ($\alpha$, $\beta$, $\nu$, the system size $L_x\times L_y$) although the actual transition to turbulence follows one of two standard routes. In one case we find that turbulence emerges through a sequence of several Hopf bifurcations, commonly referred to as the Ruelle-Takens-Newhouse scenario \cite{ruelle1971,newhouse1978}. In the other, a sub-critical bifurcation leads to intermittency, as in the Pomeau-Mannevile scenario \cite{pomeau1980}.

The details of the bifurcation sequence, however, are quite important in describing the evolution of the transport properties of the flow. As a general trend, we find the mixing efficiency (defined either in terms of the mixed area fraction or in terms of the mixing rate) to improve as the forcing is increased, with steady flows being the worst mixers and turbulent flows -- the best. However, the complexity of the flow does not increase monotonically and neither does mixing efficiency. Neither is mixing efficiency directly related to the complexity of the flow, as the comparison of three different time-periodic flows showed. Furthermore, time-periodic flows such as $P_3$ can rival the mixing efficiency of turbulent flows.

The most unexpected result was that the mixed area fraction of a class of time-periodic and quasi-periodic flow can be described -- rather accurately -- by a perturbative approach. This is despite the fact that none of the flows considered can actually be considered weakly perturbed. The description is based on the idea of multiple overlapping resonances, which define well-mixed regions of the flow. In particular, our results confirm the idea of Soskin and Mannella \cite{soskin2009} that resonances play an important role in defining the width of (and dynamics inside) the separatrix chaotic layer. Although the flow domain is densely covered by an infinite number of resonant tori, only the dominant resonances (e.g., ones that correspond to the harmonics of the frequency of the perturbation) play an important role. As a general rule of thumb, we find the flows with the broadest Fourier spectrum to possess the best mixing properties, while (nearly) monochromatic flows have mixing properties comparable to those of steady flows.

\section*{Acknowledgements}

This material is based upon work supported in part by the National Science Foundation under Grants No. CBET-0900018 and CMMI-1234436.

\bibliography{bibfile}

\end{document}